\documentclass[a4paper,11pt]{article}
\pdfoutput=1 % if your are submitting a pdflatex (i.e. if you have
\usepackage{jheppub} % for details on the use of the package, please
\usepackage{orcidlink}

\usepackage[T1]{fontenc} % Works for Umlaute
\usepackage[utf8]{inputenc} % Input
\usepackage{microtype} % Subliminal refinements towards typographical perfection
\usepackage{lmodern} % Better Font
\usepackage{booktabs} % Nicer tables
\usepackage{mdframed} % Framed boxes \begin{mdframed}
\usepackage{graphicx}

\usepackage{hyperref}
 \usepackage{xcolor}
 \definecolor{dark-red}{rgb}{0.4,0.15,0.15}
 \definecolor{dark-blue}{rgb}{0.15,0.15,0.4}
 \definecolor{medium-blue}{rgb}{0,0,0.5}
 \hypersetup{colorlinks, linkcolor={dark-red}, citecolor={dark-blue}, urlcolor={medium-blue}}

\usepackage{amsmath,amsfonts,amssymb}
\usepackage{eucal} % Nicer \mathcal{}
\usepackage{mleftright} \mleftright
\usepackage{tensor} % Tensor with \indices{^ab_cd}
\usepackage{braket} % Braket works \bra \ket \braket{ | }
\usepackage{bm} % Bold math
\usepackage{mathtools} % Useful toosl for AMS Math
\usepackage{comment}
\usepackage{mathrsfs} % Other fonts for Lagrangian, Hamiltonian, Path intgralmeasure \mathscr{LHD}

\providecommand*{\pd}{\partial}
\renewcommand*{\pd}{\partial}

\usepackage{empheq}

\title{\boldmath BMS charge algebras from\\
  Carrollian energy–momentum commutators}

\author[a,d]{Florian Ecker \orcidlink{0000-0002-0449-0081},}
\author[a]{Luciano Montecchio \orcidlink{0000-0003-2996-069X},}
\author[b]{Stefan Prohazka \orcidlink{0000-0002-3925-3983},}
\author[a,c]{Jakob Salzer \orcidlink{0000-0002-9560-344X}}

\affiliation[a]{Institute for Theoretical Physics, TU Wien, 
Wiedner Hauptstrasse 8–10/136, A-1040 Vienna, Austria
}
\affiliation[b]{University of Vienna, Faculty of Physics, Mathematical Physics, Boltzmanngasse 5, 1090, Vienna, Austria}
\affiliation[c]{Universit\'e Libre de Bruxelles and International Solvay Institutes, ULB-Campus Plaine CP231, 1050 Brussels, Belgium}
\affiliation[d]{Quantum Technology Laboratories GmbH, 
Clemens-Holzmeister-Strasse 6/6, A-1100 Vienna, Austria
}

\emailAdd{fecker@hep.itp.tuwien.ac.at} \emailAdd{luciano.montecchio@tuwien.ac.at} \emailAdd{stefan.prohazka@univie.ac.at} 
\emailAdd{jakob.salzer@tuwien.ac.at}

\abstract{Starting from background-field Ward identities, we derive
  universal local commutator relations for the (conformal) Carrollian
  energy–momentum tensor and determine conditions under which they
  reproduce the BMS charge algebra. The resulting relations separate
  universal symmetry-constrained terms from anomaly contributions and
  theory-dependent terms.  In two boundary dimensions,
  variations of a Carroll boost anomaly generate the mixed BMS$_3$
  central extension.  In three boundary dimensions, corresponding to radiative four-dimensional gravity in the bulk, we
identify local response conditions to reproduce the gravitational BMS$_4$ charge algebra. We further
  recover the Bondi flux-balance laws and the field-dependent co-cycle using a one-point holographic
  dictionary. Finally, we propose a simple Carrollian field theory action which couples to the shear.}

\begin{document} 
\maketitle
\flushbottom

\section{Introduction}
\label{sec:introduction}

Gravitational waves carry energy and angular momentum to future null infinity. Balance laws relate these outgoing fluxes to changes in the corresponding charges of the radiating
system~\cite{Bondi:1962px,Sachs:1962wk}. These charges are part of a larger set associated with the asymptotic symmetries of asymptotically flat spacetimes, which form the Bondi--van der Burg--Metzner--Sachs (BMS) algebra. Deriving the charges, their algebra, and the balance laws from an intrinsic boundary description is an important ingredient for establishing a holographic correspondence in asymptotically flat spacetimes.

The geometry of null infinity, where all outgoing gravitational radiation is being recorded, is intrinsically Carrollian. Its conformal symmetries realize the BMS
algebra~\cite{Geroch:1977big, Ashtekar:1981hw, Duval:2014uva}. It is therefore natural to look for an intrinsically Carrollian formulation of the aforementioned balance laws and charge algebra. Given the absence of a concrete microscopic Carrollian holographic dual theory, such a formulation can only refer to the universal properties of the Carrollian field theory.

In analogy to the relativistic case, coupling a field theory to Carrollian geometric background data defines response functions that obey Ward identities. 
These response functions comprise a Carrollian energy-momentum tensor (EMT), as a response to the Carroll metric data, as well as a ``news'' tensor as the response to an independent part of the Carroll connection data. The latter can be identified with the asymptotic shear~\cite{Geroch:1977big, Ashtekar:1981hw} of the asymptotically flat spacetime in which the Carrollian manifold is embedded as future null infinity.
Given a holographic dictionary, i.e., a relation between bulk quantities and response functions, one can indeed show that the boundary Ward identities reproduce the flux-balance laws~\cite{Fiorucci:2025twa, Hartong:2025jpp, Hartong:2026rbr}. 

In the present work, we shall show that the same method can be used to determine the universal part of the commutation relations of the response functions, i.e., the part of the commutation relations that is independent of the details of the theory. Integrating these commutation relations leads then directly to the algebra of charges. 

The analogous question in relativistic quantum field theory was
studied by Dirac~\cite{RevModPhys.34.592} and Schwinger~\cite{Schwinger:1959xd,Schwinger:1963zza}. The energy-density commutator provides a local condition for Lorentz invariance, which Schwinger~\cite{Schwinger:1963zza} described as \emph{``what may well be considered the most fundamental equation of relativistic quantum field theory.''} Closely related results have been obtained from a Hamiltonian point of view in~\cite{Teitelboim:1972vw},
including the zero-signature case underlying Carrollian
geometry~\cite{10.1007/3-540-09238-2_90,Henneaux:1979vn}.

In our derivation, we follow the background field method introduced by Boulware and Deser~\cite{Deser:1967zzf}. We couple the theory to geometric sources and derive the Ward identities for diffeomorphisms, Weyl transformations and Carroll boosts, allowing for a boost anomaly. Varying these identities with respect to the sources gives relations for two-point functions from which the equal-time commutators can be extracted. A complication compared to the Lorentzian case follows from the presence of the shear as additional geometric background structure that leads to non-conservation of the energy-momentum tensor. 

While we provide general relations that allow for curved Carrollian backgrounds, it is instructive to display their structure in the simple setting of a flat Carrollian background. Assuming that there is no shear coupling and the Carroll boost Ward identity is non-anomalous, the five classes of equal-time commutators involving at least one energy or momentum density take the form
\begin{subequations}\label{introbrackets}
  \begin{align}
\label{intro:full-ee}
    i[\mathbf{T}^0{}_0 (x),\mathbf{T}^0{}_0 (y)]&=0    \\
    \label{intro:full-ep}
    i[\mathbf{T}^0{}_0 (x),\mathbf{T}^0{}_i (y)]&= \big(\mathbf{T}^0 {}_0 (y) \delta^k_i -\mathbf{T}^k {}_i (x)\big) \partial _k \delta ^{d}(x-y) \\
     \label{intro:full-pp}
    i[\mathbf{T}^0{}_i (x),\mathbf{T}^0{}_k (y)]&=   \big( \mathbf{T}^0 {}_k (x) \partial_i + \mathbf{T}^0 {}_i (y) \partial_k \big) \delta^{d}(x-y) + \tau^0 {}_i {}^0 {}_k(x,y)\\
     \label{intro:full-es}
    i[\mathbf{T}^0{}_0 (x),\mathbf{T}^i{}_j (y)]&= -\partial _0 \mathbf{T}^i {}_j (x)   \delta ^{d}(x-y)\\
      \label{intro:full-ps}
    i[\mathbf{T}^0{}_i (x),\mathbf{T}^j{}_k (y)]&= \big( \mathbf{T}^j {}_k (x)   \partial_i + \mathbf{T}^n{}_i(y) \, \delta_{nk} \partial^j + \mathbf{T}^j {}_i (y)   \partial_k \big) \delta ^{d}(x-y)
     + \tau^0 {}_i {}^{j }{}_k(x,y).
  \end{align}
\end{subequations}
The distributions $\tau^0 {}_i {}^{j }{}_k(x,y)$ are the remaining
Schwinger contact terms that depend on the specific theory. In
particular, \eqref{intro:full-ee} displays a defining Carrollian
feature: the vanishing of the energy--energy bracket, as emphasized in
\cite{Henneaux:2021yzg}. As we will show, this feature is lost in a theory with shear coupling, in which case the bracket is replaced by
\begin{equation}
    i[\mathbf{T}^0{}_0 (x^k),\mathbf{T}^0{}_0(y^k)] = - \left(\partial_j\Psi^{ij}(x^k) +  \partial_j\Psi^{ij}(y^k)\right) \partial_i  \delta ^2(x-y) + \tau^u{}_u{}^u{}_u(x^k,y^k).
\end{equation}
where $\Psi^{ij}$ is the response to the shear. 

The failure of the EMT brackets to close on themselves can be traced back to the theory being sourced in the presence of a generic shear coupling. Taking into account the transformation of this source by an appropriate choice of modified brackets~\cite{Barnich:2011mi}, one nevertheless recovers the BMS algebra from these brackets. Using the holographic dictionary provided in \cite{Hartong:2025jpp} and the assumption of anomalous Carroll boost Ward identities will allow us to also recover the (field-dependent) cocycles of the BMS algebra in three and four bulk dimensions. 

The paper is organized as follows. Section~\ref{sec:carr-geom} reviews
Carrollian geometry and its conformal
symmetries. Section~\ref{sec:conf-field-theor} derives the Ward
identities and the general equal-time EMT brackets from
background variations. Section~\ref{sec:univ-stress-tens} specializes
these results to flat Carrollian backgrounds with no shear coupling and discusses the realization of the Carroll and BMS$_{3}$
algebras. In Section~\ref{sec:repr-bms_4-algebra} we include the coupling to shear, derive the universal EMT brackets, show that they lead to a realization of the $\textrm{BMS}_4$ algebra using modified brackets, and discuss how a specific holographic dictionary leads to the field-dependent cocycle in the BMS algebra. In Section~\ref{sec:discussion}, we summarize our findings, provide comments on interesting open questions and provide a toy example of a Carrollian theory that couples to the shear. Appendix~\ref{sec:cons-with-carr} shows the consistency of \eqref{introbrackets} with the Carroll algebra.

\section{Carrollian geometries}
\label{sec:carr-geom}

In this section, we review the relevant aspects of conformal
Carrollian geometries, including ambiguities in the Carroll
connection and their relation to the asymptotic shear at null infinity. We introduce the background fields $(\tau_\mu,h_{\mu\nu},C_{\mu\nu})$ to
which our theories couple.

\subsection{Basic definitions}
\label{sec:basic-definitions}

We start by setting up some formalism for describing the background geometries that our field theories of interest couple to. Consider a $(d+1)$-dimensional manifold $\mathcal{M}$ with coordinates $x=(x^{0},x^{k})=(u,x^{k})$ endowed with the data $(h_{\mu\nu}, v^{\mu}, \nabla)$, where the symmetric Carroll metric $h_{\mu\nu}$ is degenerate, its kernel being spanned by the vector field $v^{\mu}$, so that $v^{\mu} h_{\mu\nu} = 0 $ and where $\nabla$ is a torsion-free connection. The inverse data, namely the
degenerate co-metric $h^{\mu\nu}$ and the Carroll one-form $\tau_{\mu}$, are introduced through the relations
\begin{align}\label{tau h}
    v^\mu \tau _\mu &=-1 & h^{\mu \nu }h_{\nu \rho }-v^\mu \tau _\rho &=\delta ^\mu _\rho & \tau _\mu h^{\mu \nu }&=0
\end{align}
and we shall write the projector onto the transverse space as $h^\alpha _\beta :=h^{\alpha \nu }h_{\nu \beta }$. We note, however, that the relations \eqref{tau h} only determine those fields up to the transformations
\begin{align}
    \delta _\lambda v^\mu &=0 & \delta _\lambda h^{\mu \nu }&=-2v^{(\mu} h^{\nu) \rho }\lambda _\rho & \delta _\lambda \tau _\mu &=-\lambda _\mu & \delta _\lambda h_{\mu \nu }&=0  \, ,
\end{align}
where $\lambda_\mu(x)$ is an arbitrary spacetime dependent one-form
satisfying $v^\mu\lambda_\mu=0$.  In a gauge theoretic setting these
are local Carroll boosts \cite{Hartong:2015xda}. In what follows we
take the variables $h_{\mu \nu }$ and $\tau _\mu $ as the independent
metric data since they allow constructing the fields with upper
indices without ambiguity. We shall also need a few derived objects
from the metric data, the Carroll acceleration $a_\mu $ as well as the
extrinsic curvature $K_{\mu \nu }$ defined by
\begin{equation}
    a_{\mu} := \mathcal{L}_{v}\tau_{\mu} = 2v^\nu \partial _{[\nu }\tau _{\mu ]}
    \qquad \qquad 
    K_{\mu\nu}:= -\tfrac{1}{2}\,\mathcal{L}_{v}h_{\mu\nu} ~.
\end{equation}
Both these objects are transverse, $v^\mu a_\mu=0= v^\nu K_{\mu \nu}$. Our choice of torsion-free affine connection $\nabla $ follows \cite{Hartong:2025jpp,Hartong:2026rbr} and is given in terms of the coefficients
\begin{align}\label{eq:connection}
    \Gamma ^\rho {}_{\mu \nu }=-v^\rho \partial _{(\mu }\tau _{\nu )}-v^\rho \tau _{(\mu }\mathcal{L}_v \tau _{\nu )}+\frac{1}{2}h^{\rho \lambda }\big(\partial _\mu h_{\lambda \nu }+\partial _\nu h_{\lambda \mu }-\partial _\lambda h_{\mu \nu }\big) ~.
\end{align}
Notably, this is not metric compatible in the sense that it does not annihilate the Carroll boost invariant data,
\begin{align}
\label{eq:nonmetricity}
    \nabla _\mu h_{\alpha \beta }=-2 \tau_{(\alpha}K_{\beta)\mu} && \nabla _\mu v^\nu =-K_\mu {}^\nu ~.
\end{align}
In fact, one may show that only through constraining the Carroll metric data by $K_{\mu \nu}=0$ there exists a Carroll connection that is both torsion free and metric compatible. In this sense the above choice of connection has ``minimal'' non-metricity since the right hand sides of \eqref{eq:nonmetricity} are only sourced by $K_{\mu \nu }$. 

We define a volume form on $\mathcal{M}$ by
\begin{align}
    e=\sqrt{\det (\tau _\mu \tau _\nu +h_{\mu \nu })}
\end{align}
and from a generic infinitesimal variation
\begin{align}
    \delta e=e\Big(-v^\mu \delta \tau _\mu +\frac{1}{2}h^{\mu \nu }\delta h_{\mu \nu }\Big)
\end{align}
it becomes evident that it is Carroll boost invariant, $\delta _\lambda e=0$. Furthermore, the identity $ \Gamma^{\alpha}{}_{\alpha \mu} = \partial_{\mu} \log e$ allows to express divergences of a vector field $B^\mu$ as $e\nabla _\mu B^\mu =\partial _\mu (eB^\mu )$. We shall consider a generalization of the covariant derivative for tensor densities of weight $w$, i.e., quantities that transform under general coordinate transformations as
\begin{align}
    \mathfrak{B}'^{\mu }_\nu =\Big\vert \frac{\partial x}{\partial x'}\Big\vert ^w\frac{\partial x'{}^\mu }{\partial x^\alpha }\frac{\partial x^\beta  }{\partial x'{}^\nu }\mathfrak{B}^\alpha _\beta ~,
\end{align}
where we use Fraktur font to denote tensor densities. In that case, the covariant and Lie derivatives are defined by
\begin{align}\label{eq:density_cov}
    \nabla _\mu \mathfrak{B}^\alpha _\beta &=\partial _\mu \mathfrak{B}^\alpha _\beta +\Gamma ^\alpha {}_{\mu \sigma }\mathfrak{B}^\sigma _\beta -\Gamma ^\sigma {}_{\mu \beta }\mathfrak{B}^\alpha _\sigma -w(\partial _\mu \log e) \mathfrak{B}^\alpha _\beta\\
    \label{eq:density_lie}
    \mathcal{L}_\xi \mathfrak{B}^\alpha _\beta &= \xi^\mu \partial_\mu \mathfrak{B}^\alpha _\beta - \mathfrak{B}^\mu _\beta \partial_\mu \xi^\alpha + \mathfrak{B}^\alpha _\mu \partial_\beta \xi^\mu + w (\partial_\mu \xi^\mu) \mathfrak{B}^\alpha _\beta
\end{align}
and likewise for more indices (by setting $w=0$, they just reduce to the usual tensorial derivatives). In our formulation, the volume form is defined as a densitized scalar of weight $w_e=1$, such that $\nabla_\mu e =0$. 

We are ultimately interested in studying conformal Carrollian field theories, where we are only provided with a conformal structure on $\mathcal{M}$ such that $(v^\mu ,h_{\mu \nu })\sim (\Omega ^{-1}v^\mu,\Omega ^2h_{\mu \nu }) $. We define isotropic infinitesimal Weyl transformations by
\begin{align}\label{Weyl}
    \delta _\sigma v^\mu &=-\sigma v^\mu &
    \delta _\sigma h_{\mu \nu }&=2\sigma h_{\mu \nu }\\
    \nonumber
    \delta _\sigma \tau_\mu &=\sigma \tau_\mu &
    \delta _\sigma h^{\mu \nu }&=-2\sigma h^{\mu \nu } 
\end{align}
for some arbitrary function $\sigma(x)$, where the weights of $\tau_\mu$ and $h^{\mu\nu}$ follow from \eqref{tau h}.

\subsection{Ambiguities in Carrollian connections}

Let us discuss the choice of our connection \eqref{eq:connection} in more detail. It is well known that the analogous statement of the uniqueness of the Levi-Civita connection does not hold for Carrollian geometries. Given some generic one-form $k_\mu $ any two connections satisfy 
\begin{equation}
    (\hat{\nabla}_\mu-\nabla_\mu)k_\nu=-\Xi^{\lambda}_{\mu\nu} k_\lambda \, ,
\end{equation}
with $\Xi^{\lambda}_{\mu\nu}=\Xi^{\lambda}_{(\mu\nu)}$ due to torsionlessness. If both of the connections are required to be structure-compatible, then
\begin{equation}
    \Xi^{\lambda}_{\mu\nu}=v^\lambda \Xi_{\mu\nu} \qquad \quad \Xi_{\mu\nu}=\Xi_{(\mu\nu)}\qquad \quad v^\mu \Xi_{\mu\nu}=0 ~,
\end{equation}
which holds for arbitrary but fixed non-metricity, in particular for the choice \eqref{eq:nonmetricity}.
In \cite{Geroch:1977big,Ashtekar:1981hw,Ashtekar:2014zsa}, this freedom in the choice of connection on null infinity is explicitly related to the asymptotic shear and therefore of physical significance. More precisely, in $d=2$ the two degrees of freedom of the traceless part of $\Xi_{\mu\nu}$ are identified with the two helicity components of the asymptotic shear $C_{\mu\nu}$.

Before proceeding, let us fix some notation. We denote the symmetric transverse trace-free components of a tensor by enclosing its indices in brackets $\langle\dots\rangle$, and we project onto them with 
\begin{equation}
    \Xi_{\langle\mu\nu\rangle} = h_{\langle\mu}^{\,\alpha} h_{\nu\rangle}^{\beta}\,
    \Xi_{\alpha\beta}
    \qquad \quad
    h_{\langle\mu}^{\,\alpha} h_{\nu\rangle}^{\beta}
    = h_{(\mu}^{\,\alpha} h_{\nu)}^{\beta}
    - \frac{h_{\mu\nu}}{d}\, h^{\alpha\beta}\,,
\end{equation}
where transversality guarantees that this is independent of the choice of inverse metric.

In the current set-up, the previous analysis suggests that we can identify $\Xi _{\langle \mu \nu \rangle }\equiv\frac12 C_{\mu \nu }$ and that the connection 
\begin{equation}\label{eq:general_conn}
    \hat{\Gamma} ^\rho {}_{\mu \nu }=\Gamma ^\rho {}_{\mu \nu }+v^\rho \left(\frac12 C_{\mu\nu}+\frac{1}{d}h_{\mu\nu} \Xi ^\alpha {}_\alpha\right) \qquad \qquad C_{\mu\nu}=C_{\langle\mu\nu\rangle}
\end{equation}
is compatible with the same non-metricity requirements \eqref{eq:nonmetricity}. By construction $v^\rho C_{\mu\nu}$ is part of the difference between two connections so that $C_{\mu\nu}$ transforms like a tensor under the Lie derivative. In $d = 1$ any symmetric transverse trace-free tensor vanishes identically, so that $C_{\mu\nu} = 0$ and the shear plays no role in that case. 

It is important to note that although the defining relations \eqref{eq:nonmetricity} are covariant under local Carroll boosts, the reference connection \eqref{eq:connection} is not invariant itself. Its transformation under a Carroll boost can be worked out as (see also Appendix B in \cite{Hartong:2026rbr})
\begin{equation}\label{tfGamma}
     \delta _\lambda \Gamma ^\alpha {}_{ \mu \nu }=v^\alpha \nabla_{(\mu} \lambda_{\nu)} + v^\alpha a_{(\mu} \lambda_{\nu)}+v^\alpha \tau_{(\mu} \mathcal{L}_v \lambda_{\nu)} - \lambda^\alpha K_{\mu \nu} ~.
\end{equation}
Note also that the transformation laws of the shear $C_{\mu\nu}$ under Carroll boosts and Weyl transformations are given by
\begin{align}\label{boostC}
\delta_{\lambda}C_{\mu\nu}&= h_{\langle \mu }^{\,\alpha } h_{\nu \rangle }^{\beta}\delta_{\lambda}C_{\alpha \beta} =-2 h_{\langle \mu }^{\,\alpha }h_{\nu \rangle }^{\beta } \big(\nabla_\alpha+a_\alpha)\lambda_\beta\\
\delta _\sigma C_{\mu \nu }&=\sigma C_{\mu \nu }~.
\end{align}
A systematic bulk derivation of this relation may be found in \cite{Hartong:2025jpp,Hartong:2026rbr}. From a gauging perspective in which the shear is considered part of the spin connection one arrives at the same transformation law \cite{Hartong:2026bvh}, see also \cite{Nguyen:2020hot, Fiorucci:2025twa}. In the present setup, however, the Carroll boost transformation law is obtained from an intrinsic boundary perspective by demanding
\begin{align}
\label{eq:condition}
   h_{\mu }^{\,\alpha } h_{\nu  }^{\beta} \delta_\lambda \left(\hat{\nabla}_\alpha \tau_\beta \right) \overset{!}{=} -h_{\mu }^{\,\alpha } h_{\nu}^{\beta}\hat{\nabla}_{\alpha} \lambda_{\beta} \quad \Longrightarrow \quad h_{ \mu }^{\,\alpha } h_{\nu  }^{\beta}\tau_\rho \delta _\lambda \hat{\Gamma} ^\rho {}_{\alpha \beta}=0 \, .
\end{align}
Imposing this condition fixes the boost transformation of the full transverse tensor $\Xi_{\mu\nu}$, whose trace-free part reproduces \eqref{boostC}. As a consequence, $\hat{\Gamma}^\mu {}_{\alpha \beta}$ becomes boost invariant whenever $K_{\mu\nu}=0$.

\subsection{Conformal Carrollian backgrounds and Bondi frames}
\label{sec:BMS}

Packing diffeomorphisms $\xi^\mu$, Carroll boosts $\lambda_\mu$ and Weyl transformations $\sigma$ into the combined variation $\delta_{\xi,\lambda,\sigma} = \delta_\xi + \delta_\lambda + \delta_\sigma$ we may write a general transformation of the Carrollian background data as
\begin{subequations}\label{transformations}
\begin{align}
    \delta_{\xi,\lambda,\sigma}  \tau _\mu &= \mathcal{L}_\xi \tau _\mu -\lambda _\mu +\sigma \tau _\mu \\
    \delta_{\xi,\lambda,\sigma} h_{\mu \nu }&=\mathcal{L}_\xi h_{\mu \nu }+2\sigma h_{\mu \nu }\\
    \label{TfC}
    \delta_{\xi,\lambda,\sigma} C_{\mu \nu }&=\mathcal{L}_\xi C_{\mu \nu }-2h_{\langle \mu }^{\,\alpha }h_{\nu \rangle }^\beta \left(\nabla_\alpha +a_\alpha  \right)\lambda _\beta + \sigma C_{\mu \nu } \,.
\end{align}
\end{subequations}
A specific family of such backgrounds is given by fixing the metric variables to
\begin{equation}
  \label{bondi}
  \tau= du  \qquad
  \qquad h = h_{ij}(x^k)dx^i dx^j.
\end{equation}
where we introduced coordinates $x=(u,x^i)$ on $\mathcal{M}$ with
$i,j,k=1,...,d$. These choices imply that one can foliate
$\mathcal{M}$ by $u=\text{const.}$ surfaces and set the objects
$K_{\mu \nu }$ as well as $a_\mu $ to zero. Such a class of geometries
is typically referred to as Bondi frames
\cite{Campoleoni:2022wmf,Fiorucci:2025twa}.

The transformation parameters $(\xi, \lambda, \sigma)$ that preserve these have to satisfy
\begin{equation}\label{framepreserv}
  \delta_{\xi,\lambda,\sigma} \tau_\mu = 0
  \qquad \qquad
  \delta_{\xi,\lambda,\sigma} h_{\mu\nu} = 0  \, .
\end{equation}
Solving this condition allows to fix part of the parameters in terms of the frame variables: $\xi$ becomes a Carroll conformal Killing vector field 
\begin{equation}\label{pa}
\mathcal{L}_\xi v^\mu = \sigma v^\mu \qquad \qquad \mathcal{L}_\xi h_{\mu\nu} = - 2\sigma h_{\mu\nu}
\end{equation}
and, using the transversality of $\lambda_\mu$, the compensating parameters are given in terms of $\xi $ as
\begin{equation}\label{compa}
  \sigma[\xi] = v^\mu \mathcal{L}_\xi \tau_\mu 
  \qquad \qquad
  \lambda_\mu[\xi] = h^\nu{}_\mu \, \mathcal{L}_\xi \tau_\nu \, .
\end{equation}
The vector fields $\xi$ solving these equations generate the infinite
dimensional conformal Carroll algebra in $d+1$ dimensions which is
isomorphic to the BMS$_{d+2}$ algebra~\cite{Duval:2014uva}. It is
instructive to solve the conditions \eqref{pa} and \eqref{compa}
explicitly. Working in the class of Bondi frames above the Carroll
connection \eqref{eq:connection} reduces to the spatial Levi-Civita
connection of the Carroll metric, given by Christoffel symbols
$\bar{\Gamma}^k{}_{ij}$ and covariant derivative
$\bar{\nabla}_i := \nabla_i [\bar{\Gamma}]$. In this frame, the
conditions \eqref{pa} are solved by \cite{Ciambelli:2019lap}
\begin{align}\label{xi}
    \xi=\Big(f(x^i)+\frac{u}{d}\bar{\nabla}_iY^i\Big)\partial _u + Y^i(x^j)\partial _i
\end{align}
with $f$ an arbitrary function of the spatial coordinates identified as the parameter for supertranslations and $Y^i$ being constrained by
\begin{align}\label{Yprop}
    \bar{\nabla}_{(i}Y_{j)}=\frac{1}{d}h_{ij}\psi \qquad \qquad\psi = \bar{\nabla}_k Y^k ~.
\end{align}
For $d=2$ this equation leads to local conformal Killing vectors that
form the space of superrotations, while for higher dimensions the
space of $Y^i$ is only finite-dimensional. The compensating parameters
are now given by
\begin{align}
  \sigma(f,Y) =-\frac{1}{d}\psi && \lambda _\mu(f,Y) =\partial_\mu f + \frac{u}{d}\partial_\mu \psi \, .
\end{align}
Since the vector fields $\xi(f,Y)$ close under the Lie bracket, they
form a representation of the BMS$_{d+2}$ algebra
\begin{align}\label{eq:BMSLiebr}
    [\xi_1 (f_1,Y_1),\xi_2 (f_2,Y_2)]=\xi (\hat{f},\hat{Y}) \, ,
\end{align}
where 
\begin{align}
\hat{f}&=Y_1^i\partial _if_2-Y_2^i\partial _if_1 +\frac{1}{d}\big(f_1\bar{\nabla}_iY^i_2-f_2\bar{\nabla}_iY^i_1\big) & \hat{Y}^i&=Y_1^j\partial _j Y_2^i-Y_2^j\partial _jY_1^i \, .
\end{align}
Finally, note that the shear $C_{ij}$ is not left invariant by our frame-preserving transformations \eqref{framepreserv}. On the contrary, defining $\delta_{\xi(f,Y),\lambda(f,Y),\sigma(f,Y)}= \delta_{f,Y}$, it transforms according to
\begin{equation}\label{tfshear}
\delta_{f,Y}C_{ij} = \Big(\xi ^u \partial _u + \mathcal{L}_Y - \frac{1}{d}\psi\Big)  C_{ij}  - 2 \bar{\nabla}_{\langle i} \bar{\nabla}_{j \rangle} \xi ^u \, .
\end{equation}

\section{Conformal field theories on Carrollian backgrounds}
\label{sec:conf-field-theor}

Recent work has highlighted subtleties in Carrollian quantization and
continuum limits~\cite{Cotler:2024xhb}, in the vacuum
structure and representations of conformal Carrollian
theories~\cite{deBoer:2023fnj,Cotler:2024cia,Fredenhagen:2026pia}, as well as potential non-localities in holographic Carrollian theories~\cite{Cotler:2025npu}.
Motivated by these subtleties, we adopt a generating functional
approach that allows us to formulate the symmetry constraints without
choosing a particular microscopic realization.  We couple the theory
to a non-dynamical background Carrollian geometry and define its response functions through variations of the generating functional, assuming that the required functional derivatives exist. Symmetries of the theory yield Ward identities, from which we derive general relations among response functions, separating symmetry constraints from theory-dependent response terms. We shall make the following
assumptions about the Carrollian field theory:
\begin{itemize}
    \item There exists a generating functional $W[\tau_\mu,h_{\mu\nu},C_{\mu\nu}]$ whose required functional derivatives are well-defined, with the boundary-state data held fixed.
    \item The generating functional $W$ is invariant under general coordinate transformations, Weyl rescalings and, up to anomalies, local Carroll boosts.
    \item The generating functional admits a path integral realization in terms of fundamental fields $\phi$ and some matter action $S[\tau _\mu ,h_{\mu \nu }, C_{\mu \nu};\phi ]$,  which may possibly be non-local in~$\phi$. 
    \item There exists an operator representation of the response functions acting on a vector space of states, which is local in the insertion points but potentially non-locally constructed in terms of $\phi$. 
    \item  We assume
        that the response operators depend locally on the geometric
        sources, so that their variations are contact terms
        (like the last term in \eqref{thisone}). This is the case if the background fields couple locally in the action.
\end{itemize}

\subsection{Carrollian response functions}
\label{sec:response}
As our starting point, we assume there exists some generating functional $ W[\tau _\mu ,h_{\mu \nu },C_{\mu \nu }]$ of the Carroll geometric data whose variation reads
\begin{align}
  \label{var}
    \delta W[\tau _{\mu }, h_{\mu \nu },C_{\mu \nu }]=
    \int d ^{d+1}x\, e\left[\langle T^\mu \rangle \delta \tau _\mu +\frac12 \langle T^{\mu \nu }\rangle \delta h_{\mu \nu }+\frac12 \langle \Psi ^{\mu \nu } \rangle \delta C_{\mu \nu }\right]~,
\end{align}
which defines the corresponding local response functions
\begin{align}\label{eq:ummm}
  \langle T^\mu (x)\rangle &= \frac{1}{e}\frac{\delta W}{\delta \tau _\mu (x)} & \langle T^{\mu \nu }(x)\rangle &=\frac{2}{e}\frac{\delta W}{\delta h_{\mu \nu }(x)} & \langle \Psi^{\mu \nu} (x) \rangle = \frac{2}{e}\frac{\delta W}{\delta C_{\mu \nu }(x)} ~.
\end{align}
 Following \cite{Hartong:2025jpp,Hartong:2026rbr} we refer to this collection of functions as the energy-momentum-news complex (EMT-news complex). Note that, although $C_{\mu\nu} = C_{\langle\mu\nu\rangle}$ by definition, the variation $\delta C_{\mu\nu}$ can still pick up longitudinal and trace pieces. However, we may choose $\langle \Psi^{\mu\nu}\rangle $ to be symmetric transverse and trace-free without any loss of generality.\footnote{From $\delta C_{\mu\nu} = \delta\big(h_{\langle\mu}^{\alpha} h_{\nu\rangle}^{ \beta} C_{\alpha\beta}\big) = h_{\langle\mu}^{ \alpha} h_{\nu\rangle}^{\beta}\,\delta C_{\alpha\beta} + (\dots)\delta h_{\alpha\beta} + (\dots)\delta\tau_\alpha$, we see that the longitudinal and trace parts of $\delta C_{\mu\nu}$ are not independent data. Absorbing them into $\langle T^{\mu\nu}\rangle $ and $\langle T^\mu \rangle $ leaves the corresponding components of $\langle \Psi^{\mu\nu}\rangle $ undetermined, so we can set them to zero. This renders the response functions independent, i.e., a variation of $\tau_\mu$ only switches on $\langle T^\mu \rangle $ and so on. The same choice is made in \cite{Hartong:2025jpp}.}
From this viewpoint, the action couples only to the projected part of $\delta C_{\mu\nu}$, which can then be treated as independent from $\delta h_{\mu\nu}$ and $\delta\tau_\mu$. This will become useful soon when considering general variations of the EMT-news complex components. 
 
Thinking of $\exp (iW)$ as a transition amplitude between generic
states of the boundary matter theory on a fixed background, i.e.,
$\exp (iW)=\braket{\phi(u_f)|\phi(u_i)}$, we may schematically think of a
path integral realization
 \begin{align}
 \label{eq:partitionfunction}
    Z[\tau _\mu ,h_{\mu \nu }, C_{\mu \nu}]= e^{iW[\tau _\mu ,h_{\mu \nu }, C_{\mu \nu} ]}=\int _{\phi (u_i)=\phi _i} ^{\phi (u_f)=\phi _f} \mathcal{D}\phi \, e^{iS[\tau _\mu ,h_{\mu \nu }, C_{\mu \nu};\phi ]}  ~,
\end{align}
where the functional integral extends over all field configurations of $\phi $ that obey fixed boundary conditions at initial and final times $u_i$, $u_f$. We define the expectation values of arbitrary operator insertions as
\begin{align}
    \langle T^*\{ \mathcal{O}(x_{1}) \dots \mathcal{O}(x_{n})\}\rangle =\frac{1}{Z}\int _{\phi (u_i)=\phi _i} ^{\phi (u_f)=\phi _f} \mathcal{D}\phi \,\mathcal{O}(x_{1}) \dots\mathcal{O}(x_{n})\, e^{iS[\tau _\mu ,h_{\mu \nu }, C_{\mu \nu};\phi ]} ~,
\end{align}
where $T^* \{ ...\}$ denotes the covariant time-ordering with respect to a time direction singled out by the choice of background ($u$ in the above
expressions).

One may generate higher point functions by taking successive
variations with respect to the background quantities, e.g.,
\begin{align}\nonumber
     \Big(\frac{1}{e(y)}\frac{\delta }{\delta \tau _\nu (y)}\Big)\Big(\frac{1}{e(x)}\frac{\delta }{\delta \tau _\mu (x)}\Big)W&=\frac{1}{e(y)}\frac{\delta}{\delta \tau _\nu (y)}\langle T^\mu (x)\rangle \\
     &= i\langle T^* \{T^\mu (x) T^\nu (y) \}\rangle_c + \frac{1}{e(y)}\Big \langle \frac{\delta T^\mu (x)}{\delta \tau _\nu (y)}\Big \rangle \, , \label{thisone}
\end{align}
where $\langle T^* \{T^\mu (x) T^\nu (y) \}\rangle_c$ is the connected
two-point function. The terms
$\langle \delta T^\mu (x)/\delta \tau _\nu (y)\rangle $ are by assumption contact
terms, i.e., terms with support only at $x=y$. Their origin lies in
the dependence of $T^\mu (x)$ on the background geometry as well as
additional counterterms one may have added to the action $S$ by some
renormalization procedure. In the Lorentzian context they were
originally referred to as ``sea-gull'' terms see, e.g.,
\cite{Gross:1969zow,Jackiw:1985} and are essential for maintaining covariance and gauge invariance.\footnote{In an operator language, the equivalent statement is a dependence of the composite operator $T^\mu$ on the background via some regularization procedure (e.g., point-splitting by connecting with a Wilson line). This is the essence of the ``explicit'' dependence of the currents on the background in \cite{Schwinger:1963zza}.} We shall find a similar behaviour in the Carrollian setting when looking at commutators between response functions, i.e., that symmetry alone dictates a certain contact structure in the commutators.

In what follows it shall be practical to combine the responses to the
Carroll metric data into a single object \cite{Baiguera:2022lsw},
\begin{equation}
  \label{eq:SET}
\langle \mathbf{T}^{\mu}{}_\nu \rangle :=\langle T^\mu \rangle \tau_\nu +\langle T^{\mu \rho} \rangle h_{\rho \nu} \, ,
\end{equation}
and define densitized versions of the various responses by
\begin{equation}
\langle\mathfrak{T}^\mu \rangle= \langle e T^\mu\rangle \qquad \quad   \langle\mathfrak{T}^{\mu \nu}\rangle = \langle e T^{\mu \nu}\rangle \qquad \quad  \langle\mathfrak{T}^\mu{}_\nu\rangle= \langle e\mathbf{T}^\mu{}_\nu \rangle\qquad \quad \langle\mathfrak{P}^{\mu \nu}\rangle =\langle e \Psi^{\mu \nu} \rangle~.
\end{equation}
All of these carry density weight $w=1$. One may generate $\langle\mathfrak{T}^\mu {}_\nu \rangle$ by acting with the operator 
\begin{equation}
  \label{eq:Dalphbet}
\mathcal{D}^{\alpha}{}_\beta(x) := \tau_\beta(x) \frac{\delta}{\delta \tau_\alpha(x)} + 2 h_{\lambda \beta}(x) \frac{\delta}{\delta h_{\lambda \alpha}(x)}\, ,
\end{equation}
i.e., application to the effective action yields
\begin{equation}
\mathcal{D}^{\mu}{}_\nu(x) \, W = \langle \mathfrak{T}^\mu{}_\nu (x)\rangle ~.
\end{equation}
Each further application of $\mathcal{D}^\alpha{}_\beta(y)$ raises the density weight of the resulting correlation function by one unit, at the point $y$ where the operator acts. 

\subsection{Ward identities and (non-)conservation laws}
\label{sec:ward-ident-cons}

Symmetries of the background constrain the response functions. We
assume that $W$ is independently invariant under diffeomorphisms, local Carroll boosts and Weyl rescalings given in~\eqref{transformations}. Each of these invariances gives rise to a Ward identity.

We begin by specializing the general variation \eqref{var} to the
local Carroll boost transformations. The requirement
$\delta_\lambda W \overset{!}{=} 0$ leads to the condition
\begin{align}\label{boost_WI_general}
\Big[\langle T^\mu \rangle -\big(\nabla_\nu-a_\nu\big)\langle \Psi^{\mu\nu} \rangle\Big]h_{\mu\rho}=0~,
\end{align}
which we identify as the Carroll boost Ward identity. Similarly, specializing \eqref{var} to diffeomorphisms and requiring $\delta_\xi W\overset{!}{=}0$ yields the diffeomorphism Ward identity
\begin{align}\label{diffeo_WI_general}
    \frac{1}{e}\partial_{\mu} \big( e \, \langle \mathbf{T}^\mu{}_{\alpha} \rangle + e \, \langle \Psi^{\mu \nu} \rangle C_{\nu \alpha}\big) = \langle T^\mu \rangle  \partial_\alpha \tau_\mu + \frac{1}{2} \langle T^{\mu \nu} \rangle \partial_\alpha h_{\mu \nu} + \frac{1}{2} \langle \Psi^{\mu \nu} \rangle \partial_\alpha C_{\mu \nu} ~.
\end{align}
By also demanding the generating functional to be invariant under local Weyl rescalings, $\delta_\sigma W \overset{!}{=} 0$, we obtain the Weyl Ward identity
\begin{equation}
\langle \mathbf{T}^\mu{}_\mu \rangle + \frac12 \langle \Psi^{\mu \nu} \rangle C_{\mu \nu}=0 \, ,
\end{equation}
which constrains the trace of the energy-momentum tensor $\mathbf{T}^\mu{}_\nu$. We may write the various Ward identities for densitized quantities as
\begin{subequations}\label{Wardidentities}
\begin{align}\label{diff}
    \mathrm{Diff:}&\qquad \partial _\mu  \langle \mathfrak{T}^{\mu}{}_{\nu} + \mathfrak{P}^{\mu \alpha}C_{ \alpha \nu}\rangle = \langle \mathfrak{T}^{\mu} \rangle \partial_\nu \tau_\mu + \frac{1}{2} \langle \mathfrak{T}^{\mu \alpha} \rangle \partial_\nu h_{\mu \alpha} + \frac{1}{2} \langle \mathfrak{P}^{\mu \alpha} \rangle \partial_\nu C_{\mu \alpha}\\
    \label{cboosts}
    \mathrm{C. \,\,boost:}& \qquad h_{\mu}^{\nu} \Big[\langle \mathfrak{T}^\mu \rangle -\big(\nabla _\rho -a_\rho \big)\langle \mathfrak{P} ^{\rho \mu }\rangle \Big]=e\mathcal{A}^\nu \\
    \label{weyl}
    \mathrm{Weyl:}& \qquad \langle \mathfrak{T}^\mu{}_\mu  \rangle +\frac12 \langle \mathfrak{P}^{\mu \nu } \rangle C_{\mu \nu }=0 \,  ,
\end{align}
\end{subequations}
where we have allowed for a quantum anomaly $\mathcal{A}_\nu$ to the
Carroll boost Ward identity. This is motivated by the holographic analysis of three- and four-dimensional asymptotically flat Einstein gravity~\cite{Hartong:2025jpp}, where the renormalized
on-shell boundary action is not Carroll boost invariant. Such an anomaly cannot be removed by the addition of local counterterms while keeping the other symmetries intact. As a result, $W$ is not invariant under local Carroll boosts, and its variation reads
\begin{equation}
\delta _\lambda W=-\int d^{d+1}x\, e\, \lambda _\mu \, \mathcal{A}^\mu[\tau _\mu ,h_{\mu \nu }, C_{\mu \nu} ] \, .
\end{equation}

With these ingredients it is now possible to construct currents associated to Carrollian symmetries. Consider first the identity 
\begin{align}
    \frac12 \langle \mathfrak{T}^{\mu \nu }\rangle \mathcal{L}_\xi h_{\mu \nu }+\langle \mathfrak{T}^\mu \rangle \mathcal{L}_\xi \tau _\mu +\frac12 \langle \mathfrak{P} ^{\mu \nu }\rangle \mathcal{L}_\xi C_{\mu \nu }=\partial_\mu \big(\langle \mathfrak{T}^\mu{}_\alpha \rangle \xi^\alpha+ \langle \mathfrak{P}^{\mu\rho} \rangle C_{\rho\alpha} \xi^\alpha \big) \, ,
\end{align}
which follows from the diffeomorphism Ward identity contracted with a vector field $\xi^\alpha$. Next, we have for Carroll boosts
\begin{align}
   \frac12 \langle \mathfrak{T}^{\mu \nu }\rangle \delta_\lambda  h_{\mu \nu }&+ \langle \mathfrak{T}^\mu \rangle  \delta_\lambda \tau _\mu +\frac12 \langle \mathfrak{P} ^{\mu \nu }\rangle \delta_\lambda  C_{\mu \nu } =-\partial_\mu(\langle \mathfrak{P}^{\mu\nu} \rangle\lambda_\nu) - e\mathcal{A}^\mu \lambda_\mu \, ,
\end{align}
where we have again used the anomalous Carroll boost Ward identity. Adding local Weyl rescalings, we can define a Noether current
\begin{equation}
\label{eq:NOethercurrent}
    j^\mu=\left(\mathfrak{T}^\mu{}_{\alpha} + \mathfrak{P}^{\mu \nu} C_{\nu \alpha}\right) \xi^\alpha-\mathfrak{P}^{\mu\nu}\lambda_\nu
\end{equation}
such that
\begin{equation}
\label{eq:currentcons}
\boxed{    \partial_\mu \langle j^\mu \rangle =\frac12 \langle \mathfrak{T}^{\mu \nu }\rangle \delta_{\xi,\lambda,\sigma}  h_{\mu \nu }+\langle \mathfrak{T}^\mu \rangle \delta_{\xi,\lambda,\sigma}  \tau _\mu +\frac12 \langle \mathfrak{P} ^{\mu \nu }\rangle \delta_{\xi,\lambda,\sigma}   C_{\mu \nu } + e \mathcal{A}^\mu \lambda_\mu.
}
\end{equation}
We see that this current is conserved when the anomaly vanishes on a chosen background and when we consider combined transformations that leave all the background structure invariant. This case, however, is usually highly restrictive. While the equations $\delta_{\xi,\lambda,\sigma}h_{\mu\nu}=\delta_{\xi,\lambda,\sigma}\tau_\mu =0$ of Section \ref{sec:BMS} still produce an infinite number of allowed transformations, demanding $\delta_{f,Y}C_{ij}=0$ has no nontrivial solution for generic $C_{ij}$. Motivated by the relation between conformal Carrollian geometries and the physics of asymptotically flat spacetimes, we require the weaker condition of not leaving $C_{ij}$ invariant. The transformations generated by $(f,Y^i)$ then do not correspond to usual Noether symmetries since they do not only act on dynamical fields. Instead,
once the theory responds to $C_{ij}$, they only leave the generating
functional invariant if one also transforms the background field
$C_{ij}$ appropriately. While this still leads to a representation of
the symmetry algebra in terms of charge operators, it sources a
violation of their conservation law~\eqref{eq:currentcons} and one may
interpret $\langle \mathfrak{P}^{\mu \nu}\rangle $ as sourcing the
respective flux terms. This is in line with the picture of sourced
Carrollian field theories put forward in \cite{Donnay:2022wvx}. We have some comments on this point of view in the Discussion section~\ref{sec:discussion}.

Before turning to commutation relations, let us point out that \eqref{eq:currentcons} leaves the definition of the operator $j^\mu $ ambiguous since there is always freedom to add a term of the form $\pd_{\nu}K^{[\mu\nu]}$ to \eqref{eq:NOethercurrent}.  More generally, one may also include a total derivative $\partial_\mu k^\mu$ on both sides, leading to a redefinition of the current $j^\mu\rightarrow j^\mu+k^\mu$. This freedom is not sufficient to absorb all flux terms into the divergence on the left hand side of \eqref{eq:currentcons} which goes along with the fact that the right hand side consists of true physical flux. However, we shall see in  Section~\ref{sec:hologr-dict-bms_4} that once we define BMS charges as certain integrals of these currents, these ambiguities become relevant for matching with gravitational flux-balance laws.

\subsection{Commutators}
\label{sec:quant-stress-tens}
In the following we shall show that (conformal) Carroll invariance and locality of the response functions allow to determine part of the EMT-news complex commutators for any Carrollian quantum field theory. Moreover, we identify a universal sector therein that is dictated by Carrollian symmetry alone. This is along the lines of original arguments made by Schwinger and Dirac for the Lorentzian case \cite{Schwinger:1959xd,RevModPhys.34.592}.

The starting point is to consider variations of the diffeomorphism Ward identity \eqref{diff} with respect to the background quantities $\tau_\mu$ and $h_{\mu \nu}$ which is implemented by means of the operator $\mathcal{D}^{\mu}{}_\nu$ defined in~\eqref{eq:Dalphbet}. Letting it act on the expectation value of a given $\mathcal{O}(x)$ we obtain
\begin{equation}
  \label{funcderiv}
\mathcal{D}^{\mu}{}_\nu (y) \langle \mathcal{O}(x)\rangle = i \langle T^* \{ \mathcal{O}(x)\mathfrak{T}^\mu{}_\nu (y) \} \rangle_c +  \langle \mathcal{D}^{\mu}{}_\nu (y) \mathcal{O}(x)\rangle ~.
\end{equation}

We now switch from the path integral formalism to the operator picture, where the diffeomorphism and Weyl Ward identities\footnote{In the case of the Carroll boost Ward identity, the anomaly $\mathcal{A}_\nu$ could in principle be promoted to an operator as well, although the lift is not always transparent. For the purposes of this section, the identity is most naturally read at the level of expectation values.} also hold as operator identities, allowing us to drop the expectation values. Since the path integral is defined between arbitrary initial and final states, the identities are valid for every matrix element $\langle \phi_f|\dots|\phi_i\rangle$, and hence for the operators themselves. The time-ordering in \eqref{funcderiv} is fixed only for $x \neq y$; terms supported at $x = y$ can be traded with the second term on the right-hand side. In the operator picture, we will find it convenient to use another time-ordering $T$ that differs from the covariant time-ordering $T^*$ by contact terms~\cite{Jackiw:1985} that we will fix momentarily, 
\begin{align}
  \label{timeord}
    T \{ \mathcal{O}(x)\mathfrak{T}^\mu{}_\nu (y) \}=\Theta (x^0-y^0)\mathcal{O}(x)\mathfrak{T}^\mu{}_\nu (y)+\Theta (y^0-x^0)\mathfrak{T}^\mu{}_\nu (y)\mathcal{O}(x) ~.
\end{align}
To make sense of this expression, we assume a background where $\partial _{[\mu }\tau _{\nu ]}=0$ and introduce a foliation of $\mathcal{M}$ into spatial slices. We adapt coordinates $x^\mu = (x^0, x^i)$, so that the Carrollian time $x^0$ labels the $x^0 = \text{const.}$ leaves  while $x^i$ are coordinates along them. Now, by differentiating \eqref{timeord}, we can obtain equal-time commutators for the operators
\begin{equation}
\partial_\mu \left(T \{ \mathcal{O}(x)\mathfrak{T}^\mu{}_\nu (y) \} \right)= T \{ \partial_\mu \mathcal{O}(x)\mathfrak{T}^\mu{}_\nu (y) \} + \delta^0_\mu \big[\mathcal{O}(x),\mathfrak{T}^\mu{}_\nu (y) \big] \, \delta (x^0 - y^0) ~.
\end{equation}
We complete the prescription by demanding that the diffeomorphism and the boost Ward identities (when the anomaly vanishes) hold identically inside time-ordered products, i.e., $\langle T\{\mathrm{WI}(x) \mathcal{O}(y) \} \rangle=0$. This corresponds to a choice of contact terms in \eqref{funcderiv}.

With all these ingredients we may now consider a variation of the diffeomorphism Ward identity. Applying $\mathcal{D}^{\alpha}{}_\beta$ to both sides of \eqref{diff} and dropping expectation values leads to the equal-time commutators
\begin{align}\nonumber
i\delta (x^0-y^0) &[\mathfrak{T}^0{}_\nu (x),\mathfrak{T}^\alpha{}_\beta (y)] = (\mathfrak{T}^\alpha{}_{\beta} \,   \partial _\nu + \mathfrak{T}^\alpha \partial_\nu \tau_\beta + \mathfrak{T}^{\alpha \lambda} \partial_\nu h_{\lambda \beta})\delta ^{d+1}(x-y)\\
\label{bracketsgeneric}
&+\mathcal{D}^{\alpha}{}_\beta(y) \mathfrak{T}^\mu(x) \partial_\nu \tau_\mu + \frac 12\mathcal{D}^{\alpha}{}_\beta(y) \mathfrak{T}^{\mu \rho}(x) \partial_\nu h_{\mu \rho} + \frac12 P^{\mu \rho \alpha}{}_\beta (x,y) \partial_\nu C_{\mu \rho}\\
\nonumber
&- \partial_\mu \Big(S^\mu {}_\nu {}^\alpha {}_\beta(x,y)  + P^{\mu \rho \alpha}{}_\beta (x,y) C_{\rho \nu} \Big)
\end{align}
where we have defined
\begin{align}
  \label{SandP}
S^\mu {}_\nu {}^\alpha {}_\beta (x,y):= \mathcal{D}^{\alpha}{}_\beta(y) \mathfrak{T}^\mu{}_\nu (x)  \qquad \qquad
P^{\mu \nu \alpha}{}_\beta (x,y) := \mathcal{D}^{\alpha}{}_\beta(y)\mathfrak{P}^{\mu \nu}(x) \,.
\end{align}
These are bi-tensor densities with weights $w_x =1$ and $w_y =1$, with
the indices $\mu, \nu$ transforming covariantly at the point $x$ and
$\alpha, \beta$ at the point $y$. Equation \eqref{bracketsgeneric} is 
our starting point for deriving more explicit bracket relations between the energy-momentum tensor components in subsequent sections, which is always done on specific classes of background geometries. Notice that the expression holds for generic
$(\tau_\mu,h_{\mu \nu}, C_{\mu \nu})$, with the background entering
both explicitly, through its derivatives, and implicitly, through the
bi-tensor densities $S^\mu{}_\nu{}^\alpha{}_\beta$ and
$P^{\mu \rho \alpha}{}_\beta$.

Nevertheless, equation \eqref{bracketsgeneric} does not yet represent the universal part of the brackets: some terms can still be fixed by the antisymmetry of the commutators and by demanding the expression to be consistently defined on a time slice, which forbids time derivatives acting on Dirac deltas. We postpone this analysis once we restrict to specific backgrounds in the next sections.

At this stage, however, there is still some underlying structure to be unraveled. In deriving \eqref{bracketsgeneric} we have taken the variations of the background quantities to be independent, i.e.,
$\left[\delta/\delta{\tau_{\alpha}(x)},\delta/\delta{h_{\mu\nu}(y)}\right]=0$ and $\left[\mathcal{D}^{\alpha}{}_\beta(x),\delta/\delta {C_{\mu\nu}}(y)\right] = 0$, the latter implying
\begin{equation}\label{P}
P^{\mu \nu \alpha}{}_\beta (x,y) = 2 \frac{\delta \mathfrak{T}^\alpha{}_\beta(y)}{\delta C_{\mu \nu}(x)} \, .
\end{equation}
The operators $\mathcal{D}^{\mu}{}_\nu$, on the contrary, do not commute among themselves: 
\begin{equation}
\left[ \mathcal{D}^{\alpha}{}_\beta(y), \mathcal{D}^{\mu}{}_\nu(x) \right] = \delta^{d+1}(x-y) \big(\delta^\alpha_\nu \mathcal{D}^\mu{}_\beta(x) - \delta^\mu_\beta \mathcal{D}^\alpha{}_\nu(x)  \big) \, .
\end{equation}
Applying this commutator to $W$ and by dropping expectation values, it can be shown that the  coefficients need to fulfill the symmetry condition
\begin{equation}
S^\mu {}_\nu {}^\alpha {}_\beta(x,y) = S^\alpha {}_\beta {}^\mu {}_\nu (y,x) + \big(\delta^\alpha_\nu \mathfrak{T}^\mu{}_\beta(x) - \delta^\mu_\beta \mathfrak{T}^\alpha{}_\nu(x)  \big) \delta^{d+1}(x-y) 
\end{equation}
where we have assigned the weights $w_x=0$, $w_y=1$ to the delta $\delta^{d+1}(x-y)$ and $w_x=1$, $w_y=0$ to the densitized stress tensor $\mathfrak{T}^\mu{}_\nu(x)$. We define 
\begin{equation}\label{Mshift}
M^\mu {}_\nu {}^\alpha {}_\beta(x,y) := S^\mu {}_\nu {}^\alpha {}_\beta(x,y) + \delta^\mu_\beta \mathfrak{T}^\alpha{}_\nu(x) \delta^{d+1}(x-y)
\end{equation}
such that the coefficients $M^\mu {}_\nu {}^\alpha {}_\beta (x,y)$ are symmetric in the exchange of the pair of block indices $M^\mu {}_\nu {}^\alpha {}_\beta (x,y) = M^\alpha {}_\beta{}^\mu {}_\nu(y,x)$. Shifting to this definition will allow to express the brackets in a covariant and antisymmetric form once we settle on a background geometry to work on.

By varying the diffeomorphism Ward identity
with respect to $C_{\mu\nu}$ we may also extract a bracket between the other components of the EMT-news complex. Proceeding as before, this leads to the
crossed commutators between the energy-momentum tensor and
$\mathfrak{P}^{\mu \nu}$,
\begin{align}
    i\delta (x^0-y^0)[\mathfrak{T}^0{}_\nu (x)&,\mathfrak{P}^{\alpha \beta} (y)] = \mathfrak{P} ^{\alpha \beta} (x) \partial_\nu  \delta ^{d+1}(x-y) - 2 h _\nu ^{\langle \alpha}\mathfrak{P} ^{\beta \rangle \mu } (y)  \partial_\mu  \delta ^{d+1}(x-y) \nonumber \\
    \label{bracketsgeneric2}
    &+ 2 \frac{\delta \mathfrak{T} ^{\mu }(x)}{\delta C _{\alpha \beta} (y)} \partial _\nu \tau _\mu (x)+  \frac{\delta \mathfrak{T} ^{\mu \rho}(x)}{\delta C _{\alpha \beta}(y)} \partial _\nu h_{\mu \rho }(x)+  \frac{\delta \mathfrak{P} ^{\mu \rho}(x)}{\delta C _{\alpha \beta}(y)} \partial _\nu C_{\mu \rho }(x)\\
    \nonumber
    &-\partial _\mu  P^{\alpha \beta \mu}{}_\nu(y,x)  - 2 \partial _\mu \Big(  \frac{\delta \mathfrak{P}^{\mu \rho}(x)}{\delta C_{\alpha \beta} (y)} C_{\rho \nu}(x) \Big) \, .
\end{align}
Here we used transversality of $\mathfrak{P}^{\mu \nu}$ and that the variation with respect to the shear remains transverse and traceless, i.e., $\delta C_{\mu\nu}(x)/\delta C_{\alpha\beta}(y) = h^{\langle\alpha}{}_\mu h^{\beta\rangle}{}_\nu\;\delta^{d+1}(x-y)$. The remaining set of commutators $[\mathfrak{P}^{ij}(x),\mathfrak{P}^{kl}(y)]$ cannot be obtained through our current method. Any information on these brackets must therefore come from the specific theory or from the bulk in a holographic set-up.

\section{Theories without shear coupling}
\label{sec:univ-stress-tens}
In this section we consider the energy-momentum tensor brackets
\eqref{bracketsgeneric} for theories that do not couple to the degrees
of freedom in $C_{\mu \nu }$, i.e., we set
$\langle \Psi ^{\mu \nu }\rangle =0$. As discussed at the end of
Section \ref{sec:ward-ident-cons} we expect in this case that exact
conservation laws hold once the boundary transformations preserve the
Carroll metric data, with the BMS algebra realized by ordinary
commutators of the associated Noether charges.

\subsection{Universal commutators of a Carrollian EMT}

For simplicity we shall choose a flat Carrollian spacetime as our background but still work in $d+1$ dimensions such that
\begin{align}\label{flat}
    v&=-\partial_0 & \tau &=d x^0 & h&= \delta_{ij} dx^i d x^j   &e&=1~,
\end{align}
where $i = 1,\ldots,d$. This setup is mainly motivated by flat space holography in three bulk dimensions, where $C_{\mu \nu }$ has to vanish identically because of its transversality and tracelessness. The analysis below is performed in generic dimensions and does not depend on whether the theory is conformally coupled to the background or not. The only difference for the conformal case is the additional trace Ward identity $\mathbf{T}^\mu {}_\mu =0$.

It follows from \eqref{flat} that all the Christoffel symbols identically vanish and the volume density is $e=1$. To obtain the universal part of the brackets \eqref{bracketsgeneric}, we perform the shift \eqref{Mshift}, which leads to  
\begin{align}\label{flatbrackets}
     i\delta (x^0-y^0)&[\mathbf{T}^0{}_\nu (x),\mathbf{T}^\alpha{}_\beta (y)]= 
     \Big(\mathbf{T}^\alpha {}_\beta (x) \partial _\nu + \mathbf{T}^\alpha {}_\nu (y) \partial _\beta \Big) \delta ^{d+1}(x-y)-\partial _\mu  M^\mu {}_\nu {}^\alpha {}_\beta (x,y) \, .
\end{align}
For $\alpha=0$, this expression is manifestly antisymmetric under the
exchange $(\nu,x)\leftrightarrow(\beta,y)$, as required of a
commutator.  Two features, however, are not yet manifest. First,
whenever $\nu=0$ or $\beta=0$ there emerge terms involving explicit
time derivatives of the Dirac delta function, which are inconsistent
with the equal-time nature of the commutator and should cancel for any
given theory. Second, for $\alpha=i$ and $\beta=j$ the symmetry under
exchange of the spatial indices, $\mathbf{T}^{[i}{}_k \delta^{j]k}=0$,
should also be made explicit. We can use the freedom left in
$M^\mu {}_\nu {}^\alpha {}_\beta$ to implement both requirements at
once, defining a new object $t^\mu {}_\nu{}^\alpha{}_\beta$ which
removes the time derivatives acting on deltas while satisfying
\begin{equation}
t^\mu {}_\nu {}^\alpha {}_\beta (x,y) = t^\alpha {}_\beta{}^\mu {}_\nu(y,x) 
\qquad \quad t^\mu {}_\nu {}^{[i} {}_j \, \delta^{k]j} =0 \, . 
\end{equation}
The shift is performed by
\begin{align}
  \nonumber
M^\mu {}_0 {}^0 {}_0(x,y) &= -t^\mu {}_0 {}^0 {}_0 (x,y) +  2\mathbf{T}^\mu{}_0 (x)  \, \delta ^{d+1}(x-y)\\
\nonumber
M^\mu {}_0 {}^0 {}_i(x,y) &= -t^\mu {}_0 {}^0 {}_i (x,y) + \mathbf{T}^\mu {}_i (x)   \, \delta ^{d+1}(x-y)\\
\label{tshift}
M^\mu {}_0 {}^i {}_j(x,y) &= -t^\mu {}_0 {}^i {}_j (x,y) + \delta^\mu_0 \, \mathbf{T}^i{}_j (x)   \, \delta ^{d+1}(x-y)\\
\nonumber
M^l {}_i {}^j {}_k(x,y) &= -t^l {}_i {}^j {}_k (x,y) - \mathbf{T}^n{}_i(y) \, \delta_{nk} \delta^{jl} \, \delta ^{d+1}(x-y)\\
\nonumber
M^\mu {}_i {}^0 {}_j(x,y) &= -t^\mu {}_i {}^0 {}_j (x,y) \\
  M^k {}_0 {}^i {}_0(x,y) &= -t^k {}_0 {}^i {}_0 (x,y) \, . \nonumber
\end{align}
With these changes the brackets \eqref{flatbrackets} reduce to
\begin{empheq}[box=\fbox]{align}
  \label{SET}\nonumber
    i\delta (x^0-y^0)[\mathbf{T}^0{}_0 (x),\mathbf{T}^0{}_0 (y)]&=   \partial_\mu t^\mu {}_0 {}^0 {}_0 (x,y)\\
     \nonumber
    i\delta (x^0-y^0)[\mathbf{T}^0{}_0 (x),\mathbf{T}^0{}_i (y)]&= \big(\mathbf{T}^0 {}_0 (y) \delta^k_i -\mathbf{T}^k {}_i (x)\big) \partial _k \delta ^{d+1}(x-y) + \partial _\mu t^\mu {}_0 {}^{0 }{}_i(x,y)\\
    i\delta (x^0-y^0)[\mathbf{T}^0{}_i (x),\mathbf{T}^0{}_k (y)]&=   \big( \mathbf{T}^0 {}_k (x) \partial_i + \mathbf{T}^0 {}_i (y) \partial_k \big)   \delta ^{d+1}(x-y) + \partial _\mu t^\mu {}_i {}^{0 }{}_k(x,y)\\
    \nonumber
    i\delta (x^0-y^0)[\mathbf{T}^0{}_0 (x),\mathbf{T}^i{}_j (y)]&= -\partial _0 \mathbf{T}^i {}_j (x)   \delta ^{d+1}(x-y) + \partial _\mu t^\mu {}_0 {}^{i }{}_j(x,y)\\
    \nonumber
    i\delta (x^0-y^0)[\mathbf{T}^0{}_i (x),\mathbf{T}^j{}_k (y)]&= \big( \mathbf{T}^j {}_k (x)   \partial_i + \mathbf{T}^n{}_i(y) \, \delta_{nk} \delta^{jl} \partial_l + \mathbf{T}^j {}_i (y)   \partial_k \big) \delta ^{d+1}(x-y)\\
    \nonumber
    &\quad + \partial _\mu t^
\mu {}_i {}^{j }{}_k(x,y)
\end{empheq}
where the right hand sides may be seen to consist of two contributions. The terms $t^\mu {}_\nu {}^\alpha {}_\beta (x,y)$ are equal-time contact terms which are in general theory dependent and may vanish in some cases but not in others. The remaining contributions are universal terms that any Carrollian EMT must obey as a consequence of symmetry. In deriving these expressions we made use of only the Carroll boost\footnote{Throughout this analysis we assume that the boost Ward identity is either non-anomalous, or anomalous with an anomaly that vanishes identically for our flat background. As we will see, each of these cases fixes the arbitrary pieces of the brackets differently.} and diffeomorphism Ward identities as operator statements. Explicitly, the former gives the vanishing of the energy flux, i.e., $\mathbf{T}^i{}_0=0$  while the latter is just $\partial_\mu \mathbf{T}^\mu{}_\nu =0$. 

For each bracket we can integrate the $x^0$ coordinate in an infinitesimal shell around $y^0$ to get rid of the $\delta (x^0-y^0)$ factors. In doing so, we define
\begin{equation}
\tau^0{}_\nu{}^\alpha{}_\beta(y^0,x^k,y^k)= \int_{y^0-\epsilon}^{y^0+\epsilon} d x^0 \, \partial_\mu t^\mu {}_\nu {}^\alpha {}_\beta(x,y) 
\end{equation}
such that antisymmetry implies $\tau^\mu{}_\nu{}^\alpha{}_\beta(x^k,y^k) = - \tau^\alpha{}_\beta{}^\mu{}_\nu(y^k,x^k)$. We refer to these pieces as Schwinger terms. They are supported only at $x^k = y^k$ and depend on the theory. Even so, they can contribute to the brackets of integrated charges, where they produce central extensions and carry the imprint of a possible anomaly. They are not all independent; the symmetries of the theory fix some of them and relate others:
\begin{itemize}
\item The variations of the boost Ward identity \eqref{cboosts} with respect to the background data allow to fix the components $\tau^0{}_0{}^\alpha{}_\beta$ in terms of the anomaly, see~\eqref{anomalousSchwinger}. In particular, they all vanish for a non-anomalous theory. 
\item Inserting the Weyl Ward identity \eqref{weyl} into the brackets yields
\begin{align}\label{weylcondition}
\tau^0{}_0{}^\alpha{}_\alpha &= 0 & \tau^0{}_i{}^\alpha{}_\alpha &=- \mathbf{T}^k{}_i (y) \partial_k \delta^d(x-y) \, .
\end{align}
The first relation implies a consistency condition on the boost anomaly, which comes from demanding it to be Weyl covariant. The second one shows that obtaining the universal part of a Carrollian CFT requires the additional shift 
\begin{equation}
t^l {}_i {}^j {}_k \to t^l {}_i {}^j {}_k - \frac{\delta^j_k}{d} \mathbf{T}^l{}_i(y) \delta^{d+1}(x-y) \, . 
\end{equation}
\item The remaining Schwinger terms can be constrained further by demanding consistency with respect to a specific algebra of charges. In Appendix \ref{sec:cons-with-carr} we work out one such possibility, namely the algebra that follows when the finite Carroll group acts unitarily on the energy-momentum tensor.
\end{itemize}

\subsection{The Carroll boost anomaly and BMS$_{3}$}
\label{sec:boost-anomaly}

We turn our attention to the $d=1$ case where two dimensional conformally flat Carroll spacetimes serve as a natural boundary setting for flat space holography in three bulk dimensions. In that context, one expects a Carroll boost anomaly to be present \cite{Campoleoni:2022wmf,Hartong:2025jpp,Krishna:2026sba,Delfante:2026ads}, which is generically given by a local expression in terms of background quantities
\begin{align}\label{anomaly1+1}
    \mathcal{A}^\mu=\frac{c_M}{12\pi }\,h^{\mu \sigma }v^\rho \partial _{[\rho }\big(a_{\sigma ]}+K\tau _{\sigma ]}\big) \, .
\end{align}
Here, $c_M$ is some real number that plays the role of a central charge. In the flat background~\eqref{flat}, this anomaly vanishes trivially and the Carroll boost Ward identity holds without anomaly at the operator level. However, the variation of the anomaly does not vanish on this background, and it is related to some contact terms $t^\mu {}_\nu{}^\alpha{}_\beta$ by the boost Ward identity. In particular, applying $\mathcal{D}^\alpha{}_\beta$ to~\eqref{cboosts} yields
\begin{equation}\label{anomalousSchwinger}
\tau^0{}_0{}^\alpha{}_\beta(x^1,y^1)= -\int_{y^0-\epsilon}^{y^0+\epsilon} d x^0 \, \partial_k \left( \mathcal{D}^\alpha{}_\beta(y) \mathcal{A}^\mu(x)\Big\vert_{\textrm{flat}} \delta_{\mu}^k \right) \,,
\end{equation}
where the variations of \eqref{anomaly1+1} can be explicitly worked out as
\begin{align}\label{var_anomaly}
\mathcal{D}^0{}_0(y) \mathcal{A}^\mu(x)\Big\vert_{\textrm{flat}} \delta_{\mu}^1 &=  - \frac{c_M}{24 \pi} \partial_0 \partial_1 \, \delta^2(x-y) = -\mathcal{D}^1{}_1(y) \mathcal{A}^\mu(x)\Big\vert_{\textrm{flat}} \delta_{\mu}^1 \\
\mathcal{D}^0{}_1(y) \mathcal{A}^\mu(x)\Big\vert_{\textrm{flat}} \delta_{\mu}^1 &= - \frac{c_M}{24 \pi} \partial_1^2 \delta^2(x-y) \, .
\end{align}
To simplify notation we denote spatial derivatives by $\partial _1:=\partial $ and the spatial coordinate by $x^1:=\mathrm{x}$. Thus, the presence of the anomaly implies a specific contact structure in \eqref{SET} leading to
\begin{align}\nonumber
    i [\mathbf{T}^0{}_0 (\mathrm{x})&,\mathbf{T}^0{}_0 (\mathrm{y})]=0  \\
    i[\mathbf{T}^0{}_0 (\mathrm{x})&,\mathbf{T}^0{}_1 (\mathrm{y})]= \big(\mathbf{T}^0 {}_0 (\mathrm{y})  -\mathbf{T}^1 {}_1 (\mathrm{x})\big) \partial  \delta (\mathrm{x}-\mathrm{y}) +\frac{c_M}{24\pi } \partial ^3\delta (\mathrm{x}-\mathrm{y})\\
    \nonumber
     i[\mathbf{T}^0{}_1 (\mathrm{x})&,\mathbf{T}^0{}_1 (\mathrm{y})]=   \big( \mathbf{T}^0 {}_1 (\mathrm{x})  + \mathbf{T}^0 {}_1 (\mathrm{y})  \big)\partial   \delta (\mathrm{x}-\mathrm{y}) +   \tau^0 {}_1 {}^{0 }{}_1(\mathrm{x},\mathrm{y})  \, .
\end{align}
We assume our theory to be conformally coupled to the flat background such that the energy-momentum tensor fulfills the trace Ward identity $\mathbf{T}^0{}_0 = - \mathbf{T}^1{}_1$. One may construct the charges associated to these conformal symmetries in terms of the two-dimensional Carroll conformal Killing vectors $\xi$ given in \eqref{xi} and the current $j^\mu$ in \eqref{eq:NOethercurrent},
\begin{align}
    Q_{(f,Y)}= -\int d \mathrm{x}\,j^0 =  -\int d\mathrm{x}\, \Big[Y\, \mathbf{T}^0{}_1+\big(x^0\partial Y+f\big)\mathbf{T}^0{}_0\Big],
\end{align}
which are conserved, i.e., $\dot{Q}_{(f,Y)} = 0$. The EM tensor brackets imply the charge algebra
\begin{align}\nonumber
    -i[Q_{(f_1,0)},Q_{(f_2,0)}]&=0\\
    -i[Q_{(0,Y_1)},Q_{(f_2,0)}]&=Q_{(Y_1\partial f_2-f_2\partial Y_1,0)}+\frac{c_M}{24\pi}\int d \mathrm{x}\, \partial ^3Y_1f_2 \\
    \nonumber
    -i[Q_{(0,Y_1)},Q_{(0,Y_2)}]&= Q_{(0,Y_1\partial Y_2-Y_2\partial Y_1)}-\int d \mathrm{x}d \mathrm{y}\, Y_1(\mathrm{x})Y_2(\mathrm{y})\tau^0{}_1{}^0{}_1(\mathrm{x},\mathrm{y}) ~.
\end{align}
To bring this into a more recognizable form from the perspective of the Carrollian CFT literature one compactifies the spatial domain to a circle and defines Fourier modes 
\begin{align}
    M_n:=Q_{(e^{in\mathrm{x}},0)} && L_n:=Q_{(0,e^{in\mathrm{x}})}
\end{align}
such that 
\begin{equation}\label{eq:exteneded_bms3}
\begin{aligned}
    [L_n,L_m]&=(n-m)L_{n+m}\\
    [L_n,M_m]&=(n-m)M_{n+m}+\frac{c_M}{12} n^3\delta _{n+m,0}\\
    [M_n,M_m]&=0 ~.
\end{aligned}
\end{equation}
Here we set the model dependent terms $\tau^0{}_1{}^0{}_1$ to zero for simplicity. This is the centrally extended BMS$_{3}$ algebra which can be derived from a bulk perspective as the asymptotic symmetry algebra of three dimensional asymptotically flat Einstein gravity \cite{Barnich:2006av, Barnich:2014kra}, where $c_M=3/G_N$. In our derivation, we provide a boundary Ward-identity interpretation of this classical central extension.

While we have set the model-dependent terms $\tau^0{}_1{}^0{}_1$ to
zero, there are situations where they contribute in a non-trivial
way to the brackets. A notable example is the case where they are just
given by the c-numbers
\begin{align}
    \tau^0{}_1{}^0{}_1(\mathrm{x},\mathrm{y})=\frac{c_L}{24\pi }\partial ^3\delta (\mathrm{x}-\mathrm{y})
\end{align}
which leads to the typical Virasoro central term in the $[L_n,L_m]$
superrotation sector,
\begin{align}
    [L_n,L_m]&=(n-m)L_{n+m}+\frac{c_L}{12} n^3\delta _{n+m,0}~.
\end{align} 
From a bulk perspective this central charge is associated with the existence of a gravitational (diffeomorphism) anomaly \cite{Kraus:2005zm,Bagchi:2021gai} that reflects parity violation in the gravity theory. A simple example is topologically massive gravity \cite{Bagchi:2012yk}, where adding a gravitational Chern--Simons term to the Einstein–Hilbert action yields $c_L \neq 0$. From an intrinsically Carrollian context, another example is the Carroll--Ising model, see \cite{Grumiller:2025rtm} and references therein.

\section{Theories with shear coupling}
\label{sec:repr-bms_4-algebra}

In this section we turn to sourced theories, i.e., theories that couple to the degrees of freedom in $C_{\mu\nu}$ such that $\langle \Psi^{\mu\nu} \rangle \neq 0$. We work in two spatial dimensions, the case relevant for the
null boundary of four-dimensional asymptotically flat Einstein gravity, where the symmetric traceless shear $C_{\mu\nu}$ picks up two degrees of freedom. We consider commutators for the Carrollian EMT-News complex and compute its universal sector. The currents defined in \eqref{eq:currentcons} are no longer conserved and, hence, the BMS$_{4}$ algebra is not realized by canonical Poisson brackets but through modified brackets which subtract the flux contribution.

\subsection{Universal commutators of the Carrollian EMT-News complex}

For our $d=2$ setup we may consider the geometry in \cite{Barnich:2011mi} given by conformal $\mathbb{R} \times S^2$,
\begin{align}
  \label{sphere}
\tau &= du & v &= -\partial_u & h_{\mu \nu}(x^k) &= e^{2 \varphi(x^k)} \delta_\mu^{i} \delta_{\nu}^{j} \gamma_{ij}(x^k)
\end{align}
with coordinates $x=(u,x^i)$ for $i =1,2$, the metric of the spatial $2$-sphere $\gamma_{ij}$ and a scalar function $\varphi(x^k)$ that parametrizes a conformal class of metrics.  This background is adapted to the Bondi frame introduced in \eqref{bondi} where the connection $\bar{\Gamma}$ reduces to the spatial Levi-Civita one on the sphere and the curvature scalar defined by $R=h^{\mu \nu }R^\rho {}_{\mu \rho \nu }[\Gamma ]$ becomes $\bar{R}= 2e^{-2\varphi} - 2 \bar{\nabla}^2 \varphi$.

Now we turn our attention to the commutators \eqref{bracketsgeneric} evaluated on the spatial sphere \eqref{sphere}. First, we notice from the definition \eqref{SandP} that
\begin{equation}
\bar{\nabla}_\mu S^\mu {}_\nu {}^\alpha {}_\beta= \partial_\mu S^\mu {}_\nu {}^\alpha {}_\beta - \frac12 \mathcal{D}^{\alpha}{}_\beta(y) \mathfrak{T}^{\mu \rho}(x) \partial_\nu h_{\mu \rho} - (\bar{\Gamma}^\alpha_{\mu \nu} \mathfrak{T}^\mu{}_\beta + \bar{\Gamma}^\lambda_{\mu \nu} h_{\lambda \beta} \mathfrak{T}^{\mu \alpha } )\delta^{3}(x-y) \, ,
\end{equation}
keeping in mind that the covariant derivative acts on the point $x$ and, therefore, it only sees the weight $w_x=1$ and the indices $\mu,\nu$. After the shift \eqref{Mshift} and some algebra, we arrive at the brackets
\begin{align}\nonumber
&i\delta (u^x-u^y) [\mathfrak{T}^u{}_\nu (x),\mathfrak{T}^\alpha{}_\beta (y)] = \Big(\mathfrak{T}^\alpha{}_\beta(x) \partial_\nu + \mathfrak{T}^\alpha{}_\nu(y) \partial_\beta \Big)\delta^3(x-y)  - \bar{\Gamma}^\alpha_{\mu \nu} \mathfrak{T}^\mu{}_\beta \delta^3(x-y) \\
\label{generalbrackets}
& - \bar{\nabla}_\mu M^\mu {}_\nu {}^\alpha {}_\beta -\partial_\mu \big(P^{\mu \rho \alpha}{}_\beta C_{\rho \nu}\big) + \frac12 P^{\mu \rho \alpha}{}_\beta \, \partial_\nu C_{\mu \rho} \, .
\end{align}
As our ultimate goal is to relate these commutators to charge algebras defined through $\mathfrak{T}^u{}_\mu$, we will focus on brackets with $\alpha=u$. Since $\bar{\Gamma}^u_{\mu\nu}=0$ the connection piece drops in that case and, when the $P$-pieces are switched off additionally, the brackets become antisymmetric under the exchange $(\nu,x) \leftrightarrow (\beta,y)$. This shows again that the shift to $M$ is a necessary step towards establishing the universal part of the commutators. Nevertheless, there are still time derivatives of delta functions arising, which we again need to cancel by appropriate redefinitions of the $M$-pieces. We follow the shift performed in \eqref{tshift} for a flat background and generalize it to our background of interest. This leads to the redefinitions
\begin{align}\nonumber
M^\mu {}_u {}^u {}_u(x,y) &= -t^\mu {}_u {}^u {}_u (x,y) +  2\mathfrak{T}^\mu{}_u (x)  \, \delta ^{3}(x-y)\\
M^\mu {}_u {}^u {}_i(x,y) &= -t^\mu {}_u {}^u {}_i (x,y) + \left( \mathfrak{T}^\mu {}_i (x)  + \mathfrak{P}^{\mu k}(x) C_{ki}(x)\right) \, \delta ^{3}(x-y)\\
\nonumber
M^\mu {}_u {}^i {}_j(x,y) &= -t^\mu {}_u {}^i {}_j (x,y) + \delta^\mu_u \, \mathfrak{T}^i{}_j (x)   \, \delta ^{3}(x-y)\\
\nonumber
M^\mu {}_i {}^u {}_j(x,y) &= -t^\mu {}_i {}^u {}_j (x,y)
\end{align}
together with the additional shifts
\begin{align}
  \label{pshift}
P^{iju}{}_u (x,y)&= \mathfrak{P}^{ij}(x) \delta ^{3}(x-y) + p^{iju}{}_u (x,y) & P^{iju}{}_k (x,y)&= p^{iju}{}_k (x,y) ~.
\end{align}
After integrating in time and systematically applying the diffeomorphism Ward identity \eqref{diff} for operators, the universal sector of the equal-time brackets reads
\begin{align}
&i[\mathfrak{T}^u{}_u (x^k),\mathfrak{T}^u{}_u (y^k)] = - \left(\mathfrak{T}^i{}_u(x^k) +  \mathfrak{T}^i{}_u(y^k)\right) \partial_i  \delta ^2(x-y) + \tau^u{}_u{}^u{}_u(x^k,y^k)\nonumber\\
\label{SET2}
&i[\mathfrak{T}^u{}_u (x^k),\mathfrak{T}^u{}_i (y^k)] = \mathfrak{T}^u{}_u(y^k) \partial_i \delta^2(x-y) - \mathfrak{T}^j{}_i(x^k) \partial_j\delta^2(x-y) - \bar{\Gamma}^k_{ji} \mathfrak{T}^j{}_k \delta^2(x-y)\\
\nonumber
&\qquad - \mathfrak{P}^{jn}(x^k) C_{ni}(x^k) \partial_j \delta^2(x-y) - \frac12 \mathfrak{P}^{jk} \partial_i C_{jk} \delta^2(x-y) + \tau^u{}_u{}^u{}_i(x^k,y^k)\\
\nonumber
&i[\mathfrak{T}^u{}_i (x^k),\mathfrak{T}^u{}_j (y^k)] =\Big( \mathfrak{T}^u{}_j(x^k) \partial_i + \mathfrak{T}^u{}_i(y^k) \partial_j \Big)\delta^2(x-y) + \tau^u{}_i {}^u {}_j(x^k,y^k),
\end{align}
where no other Ward identity has been employed at this point. 
Here we have defined the ``sourced'' Schwinger terms as
\begin{equation}
\tau^u{}_\alpha{}^u{}_\beta = \int_{u^y-\epsilon}^{u^y+\epsilon}d u^x \left[ \bar{\nabla}_\mu t^\mu{}_\alpha{}^u{}_\beta(x,y) - \partial_i\big(p^{iju}{}_\beta(x,y) C_{j \alpha}(x) \big) + \frac12 p^{iju}{}_\beta(x,y) \partial_\alpha C_{ij}(x)\right]
\end{equation}
such that antisymmetry of the brackets under the exchange $(\alpha,x) \leftrightarrow(\beta,y)$ is realized by demanding $\tau^u{}_\alpha{}^u{}_\beta (x^k,y^k)= -\tau^u{}_\beta{}^u{}_\alpha (y^k,x^k)$. 

Using the Carroll boost Ward identity, that we assume to be non-anomalous for the moment, we then arrive at the sourced universal equal-time brackets
\begin{empheq}[box=\fbox]{align}
\nonumber
i[\mathfrak{T}^u{}_u (x^k),\mathfrak{T}^u{}_u (y^k)] &= - \left(\bar{\nabla}_j\mathfrak{P}^{ij}(x^k) +  \bar{\nabla}_j\mathfrak{P}^{ij}(y^k)\right) \partial_i  \delta ^2(x-y) + \tau^u{}_u{}^u{}_u(x^k,y^k)\\[3pt]
i[\mathfrak{T}^u{}_u (x^k),\mathfrak{T}^u{}_i (y^k)]
&= \Big(\mathfrak{T}^u{}_u(y^k)\delta_i^j
       -\mathfrak{T}^j{}_i(x^k)
       -\mathfrak{P}^{jn}(x^k)C_{ni}(x^k)\Big)
       \partial_j\delta^2(x-y) \label{eq:sourcedbox}\\
\nonumber
&\quad -\Big(\bar{\Gamma}^k_{ji}\mathfrak{T}^j{}_k
            +\frac12\mathfrak{P}^{jk}\partial_i C_{jk}\Big)
            \delta^2(x-y)
       +\tau^u{}_u{}^u{}_i(x^k,y^k)\\
\nonumber
i[\mathfrak{T}^u{}_i (x^k),\mathfrak{T}^u{}_j (y^k)] &=\Big( \mathfrak{T}^u{}_j(x^k) \partial_i + \mathfrak{T}^u{}_i(y^k) \partial_j \Big)\delta^2(x-y) + \tau^u{}_i {}^u {}_j(x^k,y^k) \, .
\end{empheq}
This is one of our main results. As a consistency check, after evaluating these brackets on a flat background and turning off $\mathfrak{P}^{ij}$, the unsourced commutators derived in \eqref{SET} can be recovered. 

Finally, we can also consider the shift \eqref{pshift} into the mixed brackets \eqref{bracketsgeneric2}. This is enough to establish their universal part, yielding
\begin{align}\label{mixed}
i[\mathfrak{T}^u{}_u (x^k),\mathfrak{P}^{ij} (y^k)] &= - \partial_u \mathfrak{P}^{ij}(x^k) \delta^2(x-y) + \rho^u{}_u{}^{ij}(x^k,y^k)\\
\nonumber
i[\mathfrak{T}^u{}_n (x^k),\mathfrak{P}^{ij} (y^k)] &= \mathfrak{P}^{ij}(x^k) \partial_n \delta^2(x-y) - 2 \delta_n^{\langle i} \mathfrak{P}^{j \rangle m}(y^k) \partial_m \delta^2(x-y) + \rho^u{}_n{}^{ij}(x^k,y^k)
\end{align}
where the $\rho^u{}_\alpha{}^{ij}(x^k,y^k)$ account for additional Schwinger terms constrained by the traceless condition on the last pair of indices.

\subsection{The BMS$_4$ charge algebra}
\label{sec:BMS4}

The universal brackets derived in \eqref{SET2} hold for any sourced Carrollian field theory on the background \eqref{sphere}. Their derivation uses only the diffeomorphism Ward identity, and not the Weyl or Carroll boost Ward identities, so they apply whether or not the theory is conformal or the boost symmetry is anomalous. To connect with flat space holography, for the remainder of this section we will restrict to conformal Carrollian theories. In particular, for this subsection, we will also assume a non-anomalous boost Ward identity, i.e. $\mathfrak{T}^i{}_u = \bar{\nabla}_j \mathfrak{P}^{ij}$. We can define charges through \eqref{eq:NOethercurrent} as
\begin{equation}
\label{eq:chargesdef}
Q_\xi(u) = -\int_{S^2} d^2x \, \xi^\mu \mathfrak{T}^{u}{}_\mu(u,x^i)  \, ,
\end{equation}
where $\xi^\mu$ is still an arbitrary vector field and we made explicit the fact that the charges can depend on time.
Computing its equal-time commutators from the universal part of \eqref{SET2} gives
\begin{align}\label{Qxi}
-i [Q_\xi, \mathfrak{T}^u{}_u] &= -\mathcal{L}_\xi \mathfrak{T}^u{}_u + \mathfrak{T}^\alpha{}_\beta \bar{\nabla}_\alpha \xi^\beta + \frac12 \mathfrak{P}^{ij} \mathcal{L}_\xi C_{ij}\\
\nonumber
-i [Q_\xi, \mathfrak{T}^u{}_i] &= -\mathcal{L}_\xi \mathfrak{T}^u{}_i + \mathfrak{P}^{jk} C_{ki} \partial_j \xi^u \, ,
\end{align}
where $\mathcal L_\xi$ acts on $\mathfrak{T}^\alpha{}_\beta$ as a mixed tensor density of weight $w=1$ and Schwinger term contributions were omitted for concreteness. The remaining terms in the right-hand side account for the non-trivial response of the background $(h_{\mu \nu },\tau _\mu ,C_{\mu \nu })$ under the action of $\xi$. In particular, note that
\begin{equation}
\mathfrak{T}^\alpha{}_\beta \bar{\nabla}_\alpha \xi^\beta=\frac12\mathfrak{T}^{\mu\nu} \mathcal{L}_\xi h_{\mu\nu} + \mathfrak{T}^\mu \mathcal{L}_\xi \tau_\mu \, .
\end{equation}
Since $Q_\xi$ is defined on a single slice of constant $u$, the dependence on $\partial_u\xi^\mu$ cancels out in \eqref{Qxi}. In the same vein, $u$-derivatives of $\mathfrak{T}^u{}_\nu$ can be related to purely spatial terms using the Ward identity. For the mixed brackets \eqref{mixed}, the universal part does not fix the full diffeomorphism action on $\mathfrak{P}^{ij}$, 
\begin{equation}
-i [Q_\xi, \mathfrak{P}^{ij}] = -\xi^u\partial_u\mathfrak{P}^{ij}-(\mathcal L_{\xi^k}\mathfrak{P})^{\langle ij \rangle}  \, .
\end{equation}
Instead, it only describes the flow along $\xi$ on a u-constant slice, and the remaining terms must be read off from the Schwinger terms.

The equal-time commutator of two charges is given by
\begin{equation}
\label{eq:diffeocom}
    -i[Q_{\xi_1},Q_{\xi_2}] = Q_{[\xi_1,\xi_2]}+\int_{S^2}d^2x \,\left[\mathfrak{T}^\alpha{}_\beta \left(\xi^u_1\bar{\nabla}_\alpha \xi^\beta_2-\xi^u_2\bar{\nabla}_\alpha \xi^\beta_1\right)+\frac{1}{2}\mathfrak{P}^{ij}\left(\xi^u_1\mathcal{L}_{\xi_2}C_{ij}-\xi^u_2\mathcal{L}_{\xi_1}C_{ij}\right)\right] \, .
\end{equation}
We see that these charges form the algebra of diffeomorphisms up to terms proportional to the flow of the background structure $(h_{\mu\nu},\tau_\mu,C_{\mu\nu})$ under the transformations.

As a next step, we specialize the diffeomorphisms $\xi$ to the BMS generators in \eqref{xi} such that
\begin{equation}
    \mathfrak{T}^\alpha{}_\beta \bar{\nabla}_\alpha \xi^\beta=\frac{1}{2}\psi \,\mathfrak{T}^\alpha{}_\alpha+\mathfrak{T}^i{}_u \bar{\nabla}_i\xi^u \, . 
\end{equation}
We can then impose the non-anomalous boost Ward identity as well as the Weyl Ward identity. The commutator between charges reads then
\begin{equation}
-i[Q_{\xi_1},Q_{\xi_2}] = Q_{[\xi_1,\xi_2]} 
+\frac12\int_{S^2} d^2x \, \mathfrak{P}^{ij} \left(\xi^u_1 \delta_{f_2,Y_2} C_{ij} -   \xi^u_2 \delta_{f_1,Y_1} C_{ij}\right) \, .
\end{equation}
where $\xi_1 := \xi(f_1,Y_1)$ and $[\xi_1,\xi_2]^\mu = \mathcal{L}_{\xi_1} \xi^\mu_2$. The second term in this expression corresponds to the flow piece generated by the non-trivial transformation of the shear $C_{ij}$ under BMS. In order to deal with this piece, we follow the prescription of \cite{Barnich:2011mi} and define a set of modified Barnich--Troessaert (BT) brackets for non-integrable charges
\begin{equation}
  \label{modified}
  i\left[Q_{\xi_1},Q_{\xi_2}\right]_{\mathrm{BT}}:= i\left[Q_{\xi_1},Q_{\xi_2}\right]+\delta^C_{f_1,Y_1}Q_{\xi_2}-\delta^C_{f_2,Y_2}Q_{\xi_1} \, .
\end{equation}
The variation $\delta^C$ is defined to act as
 \begin{equation}
  \label{deltaQ}
   \delta^C_{f,Y}Q=\int_{I_\epsilon\times S^2} d^{2+1}x \frac{\delta Q }{\delta C_{\mu \nu}(x)} \, \delta_{f,Y} C_{\mu \nu}(x) \, ,
 \end{equation}
 where $I_\epsilon$ is a small time-interval around the charge. For the composition of two variations acting on any functional, $[\delta^C_{f_1,Y_1}, \delta^C_{f_2,Y_2}]=-\delta^C_{\hat{f},\hat{Y}}$ holds. Using the explicit definition of the charges~\eqref{eq:chargesdef}, the identification \eqref{P} and  the shift in the contact terms $P^{iju}{}_\alpha (x,y)$ in \eqref{pshift}, we obtain
\begin{equation}
\label{eq:BMSnonco}
-i\left[Q_{\xi_1},Q_{\xi_2}\right]_{\mathrm{BT}} = Q_{[\xi_1,\xi_2]},
\end{equation}
where we neglected the Schwinger term contributions.
This corresponds to the BMS$_{4}$ algebra. We have thus shown that any field theory on a conformal Carrollian background can realize it using the BT brackets. 

If we keep track of the Schwinger terms, the above equation~\eqref{eq:BMSnonco} admits the additional piece
\begin{align}
\label{eq:BMS4Schwinger}
    K^{S}_{\xi_1,\xi_2}&=-\int_{S^2} d^2x d^2y \, \, \xi^\alpha_1 (x^k) \xi^\beta_2(y^k) \tau^u{}_\alpha{}^u{}_\beta(x^k,y^k)  \\
    \nonumber
    &\quad + \frac12\int_{I_\epsilon} du^x\int_{S^2} d^2x d^2y \big[ \xi^\alpha_2(y)  \delta_{f_1,Y_1}C_{ij}(x) - \xi^\alpha_1(y)  \delta_{f_2,Y_2}C_{ij}(x) \big]p^{iju}{}_\alpha(x,y)
\end{align}
on the right hand side. This object is an antisymmetric bilinear form in $(\xi_1,\xi_2)$ built from the Schwinger contact pieces. It can be interpreted as a possible extension of the BMS$_4$ algebra. Requiring the Jacobi identities to hold leads to the cocycle condition
\begin{equation}
\label{eq:cocyclecon}
\sum_{\textrm{cyclic}}\left(K^S_{[\xi_1,\xi_2],\xi_3} -i[K^S_{\xi_1,\xi_2},Q_{\xi_3}]_{\mathrm{BT}}\right)=0,
\end{equation}
where the action of the modified bracket is taken as
\begin{equation}
    i[K^S_{\xi_1,\xi_2},Q_{\xi_3}]_{\mathrm{BT}}=i[K^S_{\xi_1,\xi_2},Q_{\xi_3}]-\delta^C_{f_3,Y_3}K^S_{\xi_1,\xi_2}.
\end{equation}
 A cocycle of this kind appears in the asymptotic symmetry analysis of~\cite{Barnich:2011mi}. In the next subsection we show that the BT-cocycle can be reproduced even with the Schwinger terms set to zero. It follows instead from the holographic dictionary of \cite{Hartong:2025jpp} together with a Carroll boost anomaly.

\subsection{Holographic dictionary and the BMS$_{4}$ cocycle}
\label{sec:hologr-dict-bms_4}

In this subsection, we will be interested in constraining the one-point correlators of certain operators in accordance with the holographic renormalization procedure carried out in \cite{Hartong:2025jpp}. This information comes as a pure holographic input to our Carrollian CFT that otherwise would not be sensitive to the bulk data. Following the approach of that work, we enforce a local constraint on the response function in terms of the background data
\begin{equation}\label{EVpsi}
 \langle \mathfrak{P} ^{\mu\nu}\rangle = \frac12 e h^{\mu\alpha} h^{\nu \beta} \left(-\mathcal{L}_v C_{\alpha \beta} - \frac12 K C_{\alpha \beta} \right)
\end{equation}
which, evaluated on our geometry, reads
\begin{align}
\label{eq:PisN}
      \langle \mathfrak{P} ^{ij}\rangle\big \vert _{\mathbb{R}\times S^2} =\frac12 eN^{ij}, \qquad \qquad N_{ij}= \partial_u C_{ij}.
\end{align}
We identify $N_{ij}(x)$ with the Bondi news tensor, which encodes the radiative data at the asymptotic boundary in flat space holography. We also consider the result of \cite{Hartong:2025jpp} for the anomaly in the Carroll boost Ward identity. In our background this anomaly does not vanish, and instead it yields
\begin{align}\label{anomaly1+2}
    \mathcal{A}^\mu\Big \vert _{\mathbb{R}\times S^2} =h^{\mu \nu }\Big(\frac14 \partial _\nu \bar{R}+\frac12 \bar{\nabla} _\rho \partial _u C^\rho {}_\nu \Big) \, .
\end{align}
We shall show that these identifications are enough to constrain a large part of the stress tensor components through the use of the Ward identities and flux balance laws. From now on, all expressions should be understood as already evaluated in the holographic background. We start with the Carroll boost Ward identity \eqref{cboosts}, that can be worked out as
\begin{align}
\label{eq:Tiuanom}
    \langle \mathfrak{T} ^i{}_u \rangle  =e\Big(\frac14 \partial^i \bar{R} +\bar{\nabla} _kN^{ki}\Big) \, .
\end{align}
Let us stress that this identity only holds at the level of one-point expectation values. This is a sharper restriction than it may appear: in contrast with the previous section, there is no clear prescription for how it could be promoted to an operator identity. We will further comment on this formula in the discussion. Plugging \eqref{eq:Tiuanom} into the diffeomorphism Ward identity yields
\begin{subequations}\label{flux}
\begin{align}
\partial_u \langle \mathbf{T}^u{}_u \rangle &=  \frac14 N^{ij}N_{ij} - \bar{\nabla}_i\bar{\nabla}_j N^{ij}-\frac14\bar{\nabla}^k\bar{\nabla}_k \bar{R}, \\
\partial_u \langle \mathbf{T}^u{}_i \rangle &= - \bar{\nabla}_j\langle \mathbf{T}^j{}_i\rangle + \frac14 N^{jk} \bar{\nabla}_i C_{jk} - \frac12 \bar{\nabla}_j\left( N^{jk}C_{ki}\right) \, ,
\end{align}
\end{subequations}
and the Weyl Ward identity becomes
\begin{equation}
   \langle \mathfrak{T}^i{}_i\rangle + \langle \mathfrak{T}^u{}_u\rangle  + \frac{e}{4} N^{ij}C_{ij}=0 \, .
\end{equation}

As a next step, we want to define holographic BMS charges $\tilde{Q}$ at equal-time spatial slices $\Sigma$. We follow the Barnich-Troessaert prescription \cite{Barnich:2011mi}, in which the charges are written in terms of the Bondi mass $\mathcal{M}$ and the angular momentum aspect $\mathcal{N}_i$,
\begin{equation}
\langle \tilde{Q}_{(f,Y)}\rangle=\frac{1}{2\kappa}\int_{\Sigma} d^2x \, e \, \Big[4 \left( f+\frac{u}{2}\bar{\nabla}_iY^i\right)\mathcal{M} + Y^i \left(2 \mathcal{N}_i + \frac{1}{16} \partial_i( C^{jk}C_{jk}) \right) \Big] \, .
\end{equation}
The asymptotic analysis constrains these quantities through the flux-balance laws
\begin{subequations}\label{flux2}
\begin{align}
    \partial_u \mathcal{M}&=-\frac{1}{8} N_{ij}N^{ij}+\frac{1}{4}\bar{\nabla}_i\bar{\nabla}_j N^{ij}+\frac{1}{8} \bar{\nabla}^k\bar{\nabla}_k \bar{R},\\
    \partial_u \mathcal{N}_i&=\partial_i \mathcal{M}+\frac{1}{16} \partial_i\left(N^{jk}C_{jk}\right)+\frac{1}{4}C_i^{j}\partial_j \bar{R} -\frac{1}{4}\bar{\nabla}_i C^{jk}N_{jk}\\
    \nonumber
    &\quad -\frac{1}{4}\bar{\nabla}_j\left(C^{jk}N_{ki}-N^{jk}C_{ki}\right)-\frac{1}{4}\bar{\nabla}_j\left(\bar{\nabla}^j\bar{\nabla}_k C^{k}_i-\bar{\nabla}_i\bar{\nabla}_kC^{jk}\right) \, .
\end{align}
\end{subequations}
To express $\mathcal{M}$ and $\mathcal{N}_i$ in terms of our stress tensor variables, one will need to define a one-point holographic dictionary that relates these quantities. For that purpose, we match the Ward identities \eqref{flux} with the flux balance laws \eqref{flux2}, providing the identifications
\begin{subequations}\label{bondimass}
\begin{align}
2\mathcal{M} &= - \langle \mathbf{T}^u{}_u\rangle - \frac12 \bar{\nabla}_i \bar{\nabla}_j C^{ij}\\
2\mathcal{N}_i&= - 2\langle \mathbf{T}^u{}_i\rangle - \frac{1}{16}\partial_i ( C^{jk}C_{jk}) \, .
\end{align}
\end{subequations}
together with a constraint on the spatial traceless stress tensor.\footnote{For the matching of these expressions we used the identities $C^{ij}N_{jk} = - N^{ij}C_{jk} + \delta^i_k \,C^{nm}N_{nm}$ and $\left[\bar{\nabla}_i,\bar{\nabla}_j \right] \bar{\nabla}_k C^{ki} = \frac{\bar R}{2} \bar{\nabla}_k C^{k}{}_j$, which are only valid in $d=2$.
} After performing the split $\mathbf{T}^i{}_j = \hat{\mathbf{T}}^i{}_j + \frac12 h^i_j \, \mathbf{T}^k{}_k$, the constraint yields
\begin{equation}
\bar{\nabla}_j \langle \hat{\mathbf{T}}^{j}{}_i \rangle = \frac14 \bar{\nabla}_j(C^j{}_i \bar{R}) - \frac14 \bar{\nabla}_j \bar{\nabla}^j \bar{\nabla}^k C_{ki} - \frac{\bar{R}}{8} \bar{\nabla}_k C^k{}_i \, ,
 \end{equation}
 which is solved by the symmetric traceless choice
\begin{equation}
\langle \hat{\mathbf{T}}^{ij} \rangle = -\frac12 \bar{\nabla}^{(i} \bar{\nabla}_k C^{j)k} + \frac14 h^{ij} \bar{\nabla}_k \bar{\nabla}_n C^{kn} + \frac{\bar{R}}{4} C^{ij}.
\end{equation}
In this way, we have rederived the flux balance laws from the Carroll Ward identities conditional on the holographic dictionary of \cite{Hartong:2025jpp} and the existence of an anomaly. For a given Carrollian matter theory, the dictionary relates expectation values of matter operators to classical geometrical data, most importantly the shear $C_{ij}$.

Having derived the identifications \eqref{bondimass}, we may write the charges using the EMT components of the Carrollian matter theory
\begin{align}
    \tilde{Q}_{(f,Y)}=-\int_{S^2} d^2x \,\left[ \left(f + \frac{u}{2}\bar{\nabla}_iY^i \right) \left( \mathfrak{T}^u{}_u + \frac{e}{2} \bar{\nabla}_i \bar{\nabla}_j C^{ij} \right) + Y^i \mathfrak{T}^u{}_i\right], 
\end{align}
or equivalently, 
\begin{equation}\label{charges}
 \tilde{Q}_{\xi} = -\int_{S^2} d^2x \, \left( \xi^\alpha(f,Y) \, \mathfrak{T}^u{}_\alpha + \xi^u(f,Y) \frac{e}{2} \bar{\nabla}_i \bar{\nabla}_j C^{ij} \right) \, .
\end{equation}
with $\xi^\mu(f,Y)$ the generator of BMS transformations in \eqref{xi}. The normalization $\kappa=1$ was considered for simplicity. This definition of the charges matches with our Noether current \eqref{eq:NOethercurrent} by incorporating improvement terms on the EMT components: defining a vector density $k^\mu$ that parametrizes the freedom in the split of the current, such that $j^\mu \to j^\mu + k^\mu $ and
\begin{align}
    Q_{\xi}\to \tilde{Q}_{\xi}=-\int_{S^2} d^2x\, \big( j^u + k^u \big) \qquad \text{with} \qquad  j^u=\mathfrak{T}^u{}_\nu \, \xi^\nu,
\end{align}
the charges can then be reproduced by the choice 
\begin{equation}
k^\mu= \frac{e}{2} \xi^\mu \bar{\nabla}_i \bar{\nabla}_j C^{ij}.
\end{equation}
This corresponds to the addition of a c-number piece to the charge operator. Furthermore, the improved current \eqref{eq:currentcons} with anomaly \eqref{anomaly1+2} yields the non-conservation law
\begin{align}\label{nonconserved}
    \frac{d }{d u}\langle \tilde{Q}_{\xi} \rangle&= -\int_{S^2} d^2x \,  \partial_\mu\big(\langle j^\mu \rangle + k^\mu \big)\\ \nonumber
    &=-\frac{1}{4}\int_{S^2} d^2 x \, e \, \Big[N^{ij}\delta _{f,Y}C_{ij}+\partial _i\xi ^u \partial^i\bar{R}+C^{ij}\bar{\nabla}_i\bar{\nabla}_j\psi \Big].
\end{align}
The Bondi mass-loss formula can be recovered by taking $f=1,Y^i=0$. It also follows that on the sphere, where $\varphi=0$ and $\bar{R}$ is constant, the global BMS$_4$ charges are conserved in the absence of news.\footnote{When $Y^i$ are global conformal Killing vectors of the sphere, then $\bar\nabla_{\langle i}\bar\nabla_{j\rangle}\psi=0$ holds.}

We can compute the brackets between the charges using the general expression \eqref{eq:diffeocom}. However, in this case we want to  make use of the 
holographic dictionary \eqref{eq:PisN} and \eqref{anomaly1+2}, which have been 
determined only at the level of expectation values. We will omit all the Schwinger term contributions on the upcoming expressions, which amounts to setting 
$\langle K^S_{\xi_1,\xi_2} \rangle=0$ from the very start. We have
\begin{align}
  \label{eq:chargebrackalmost}
-i\langle [\tilde{Q}_{\xi_1},\tilde{Q}_{\xi_2}]\rangle  = \langle \tilde{Q}_{[\xi_1,\xi_2]}\rangle &+\frac12\int_{S^2} d^2x \, \langle \mathfrak{P}^{ij} \rangle \left(\xi^u_1 \delta_{f_2,Y_2} C_{ij} -   \xi^u_2 \delta_{f_1,Y_1} C_{ij}\right)\\
&+ \int_{S^2} d^2x \, \frac{e}{2}[\xi_1,\xi_2]^u \bar{\nabla}_i \bar{\nabla}_j C^{ij} + \int_{S^2} d^2x \, e \mathcal{A}^i\left(\xi^u_1 \partial_i \xi^u_2 - \xi^u_2 \partial_i \xi^u_1 \right) \, . \nonumber
\end{align}
The first term in the second line comes from the improvement $k^u$, while the second term derives from the anomalous contributions to the boost Ward identity. We define again modified brackets using \eqref{modified}. Due to the shift in $\tilde{Q}_\xi$ we have now
\begin{equation}
     \frac{\delta \tilde{Q}_{\xi}(u)}{\delta C_{ij}(x)}=\frac{\delta Q_{\xi}(u)}{\delta C_{ij}(x)}-\frac{e}{2}\bar{\nabla}^{\langle i}\bar{\nabla}^{j\rangle}\xi^u(x) \,\delta(u-u^x) \, .
\end{equation}
As before, the flow term in the first line of \eqref{eq:chargebrackalmost} cancels and we are left with the charge algebra
\begin{equation}
  -i\langle [\tilde{Q}_{\xi_1},\tilde{Q}_{\xi_2}]_{\mathrm{BT}}\rangle  = \langle \tilde{Q}_{[\xi_1,\xi_2]}\rangle +K_{\xi_1,\xi_2} \ ,
\end{equation}
which corresponds to the BMS$_{4}$ algebra with a field-dependent extension 
\begin{equation}
\begin{split}
     K_{\xi_1,\xi_2}= \int_{S^2} d^2x \left[\, \frac{e}{2}[\xi_1,\xi_2]^u \bar{\nabla}_i \bar{\nabla}_j C^{ij} + \,e \mathcal{A}^i\left(\xi^u_1 \partial_i \xi^u_2 - \xi^u_2 \partial_i \xi^u_1 \right)\right.\\
    \qquad\qquad \left.-\frac{e}{2} \left(\bar{\nabla}^{\langle i} \bar \nabla^{j\rangle}\xi^u_1 \delta_{f_2,Y_2} C_{ij}-\bar{\nabla}^{\langle i} \bar \nabla^{j\rangle}\xi^u_2 \delta_{f_1,Y_1} C_{ij}\right)\right].
\end{split}
\end{equation}
This piece can be further worked out: first, the contribution from the news to the anomaly cancels with a corresponding term in the second line. Then, integrating by parts and using the identity
\begin{equation}
    \left(\mathcal{L}_Y+\frac{3}{2}\psi\right)\bar{\nabla}_i\bar{\nabla}_jC^{ij}=\bar{\nabla}_i\bar{\nabla}_j\left(\mathcal{L}_Y+\frac{3}{2}\psi\right)C^{ij}+\frac{1}{2}C^{ij}\bar \nabla_i\bar \nabla_j \psi
\end{equation}
for the conformal Killing vector $Y^i$, also the first term in the first line cancels.\footnote{To derive this identity, it is useful to note that in our background $h^{ik} h^{jl} \mathcal{L}_Y C_{kl} = \mathcal{L}_Y C^{ij} + 2 \psi C^{ij}$ holds, as well as $\mathcal{L}_Y \bar{\Gamma}^k_{ki} = \partial_i \psi$. } We finally reproduce
\begin{equation}
\label{eq:BTcocycle}
K_{\xi_1,\xi_2}=\int_{S^2}d^2x\,\frac{e}{4} \Big[\big(\xi_1^u\partial_i \xi_2^u- \xi_2^u\partial_i\xi_1^u\big) \partial^i \bar R
+C^{ij} \big(\xi_1^u \bar{\nabla}_i \bar{\nabla}_j\psi_2-\xi_2^u\bar {\nabla}_i\bar{\nabla}_j\psi_1\big)\Big], 
\end{equation}
the field-dependent cocycle piece of \cite{Barnich:2011mi}.

\section{Discussion}
\label{sec:discussion}

In this work, we have discussed the analogues of the \emph{``most
  fundamental''} \cite{Schwinger:1963zza} equations of quantum field
theory in the Carrollian and conformal Carrollian case using the
background field method of \cite{Deser:1967zzf}.  We first computed
the fundamental brackets \eqref{SET} in the case when the field theory
couples only to the Carroll metric data $(\tau,h)$. This extends
the results \cite{Teitelboim:1972vw, Henneaux:2021yzg} discussed from
the Hamiltonian perspective by Schwinger terms.
We then considered the case that the theory couples in addition to the shear $C_{ij}$, showed that the fundamental brackets \eqref{eq:sourcedbox} lead to the extended BMS algebra upon using modified brackets, and reproduced the Barnich--Troessaert cocycle~\cite{Barnich:2011mi} when supplied in addition by the holographic dictionary of~\cite{Hartong:2026rbr}.
In the following, we list some interesting observations and open questions left for future investigations.
  
\begin{description}
\item[Schwinger terms.] We have been careful to keep track of
  non-universal terms in the stress tensor commutator algebra that
  depend on details of the Carrollian theory and the regularization
  scheme. In the 1+1-dimensional case we have been able to relate this
  to the presence of a Carrollian boost anomaly. It would be
  interesting to see whether explicit examples of (conformal)
  Carrollian theories can give rise to specific instances of such
  terms, even if there are known difficulties to define interacting
  Carrollian quantum field theories~\cite{Cotler:2024xhb}.

  It would be particularly interesting to see whether any field theory
  can intrinsically reproduce the Barnich--Troessaert
  cocycle~\eqref{eq:BTcocycle} from the Schwinger
  terms~\eqref{eq:BMS4Schwinger} associated with the $\textrm{BMS}_4$
  algebra, without reference to a holographic dictionary like in
  Section~\ref{sec:hologr-dict-bms_4}.
    
  \item [Holographic dictionary.] The holographic dictionary, i.e.,
    the identifications \eqref{bondimass} and \eqref{eq:PisN}, as well
    as the existence of a boost anomaly~\eqref{anomaly1+2} were
    crucial in obtaining the BT-cocycle in the extended BMS
    algebra. These relations identify expectation values of operators
    of a Carrollian theory in terms of geometric data of the
    background to which the theory couples.
    
    Although we also used a holographic dictionary for the boost anomaly in the derivation of the $\textrm{BMS}_3$-cocycle, the derivation of the $\textrm{BMS}_4$-cocycle in
    Section~\ref{sec:hologr-dict-bms_4} is on a different footing. In the
    $\textrm{BMS}_3$-case we imposed the Ward identities as operator
    identities, whereas in the $\textrm{BMS}_4$-case we had to work at
    the level of expectation values. In particular, imposing
    \eqref{EVpsi} or \eqref{anomaly1+2} as operator identities leads
    to wrong results.
    
    This suggests that there is still more to be understood about the
    holographic dictionary.  For instance, we have not been explicit
    in specifying the state in which  $\mathfrak{P}^{\mu\nu}$ is
    evaluated. It is possible that the dictionary that relates
    $\mathfrak{P}^{ij}$ to $\partial_u C_{ij}$ only holds for a
    certain class of states, e.g., coherent states that have a good
    classical description.

    However, there appears to be another issue worth
    highlighting. Taking both the definition of $\Psi^{\mu\nu}$ as a
    response to $C_{\mu\nu}$ as well as the dictionary \eqref{EVpsi}
    at face value, we have an equation of the schematic form
    \begin{equation}
    \label{eq:Cdictionary}
        \frac{\delta W}{\delta C(x)}=\langle \mathfrak P(x)\rangle\propto\partial_u C(x).
    \end{equation}
    Varying this a second time with respect to $C$ we find that the
    right hand side is anti-symmetric while the left hand side is
    symmetric. While these considerations do not contradict the
    results of~\cite{Hartong:2025jpp, Hartong:2026rbr}, they suggest
    that an additional ingredient may be needed to interpret the
    holographic response as a functional derivative of a
    boundary generating functional.

    \item[Detector operators and Carrollian operators.]
Currents on null surfaces and their algebras have attracted renewed interest
through the study of light-ray or detector operators, see, e.g., \cite{Cordova:2018ygx,Gonzo2021,Gonzalez:2025ene,Moult:2025njc, Chicherin:2025keq,Himwich:2026exq}.
At future null infinity, an energy detector measures the flux along a null
generator in a given direction \(x\). The holographic dictionary of Section \ref{sec:hologr-dict-bms_4} relates
this measurement to the Carrollian stress tensor. On the round Bondi
sphere, integrating the energy Ward identity~\eqref{flux} between two cuts gives
\begin{equation}
  \mathcal E_{\mathrm{rad}}(x)
  :=\frac14\int_{u_-}^{u_+}du\,N^{ij}N_{ij}
  =\left[
    \langle\mathbf T^u{}_u\rangle
    +\bar\nabla_i\bar\nabla_j C^{ij}
  \right]_{u_-}^{u_+} \, .
\end{equation}
It would be interesting to extend this relation to an operator identity and thereby characterize the energy detector intrinsically in terms of the Carrollian stress tensor. More generally, one may ask which detector operators at null infinity admit a similar intrinsic description in terms of Carrollian currents. We will leave this interesting direction to future work.
    
    \item [Making the background dynamical?] As we saw explicitly throughout this work, the fact that the background in the form of the shear $C_{ij}$ transforms non-trivially under BMS symmetries leads to non-conserved charges~\eqref{eq:currentcons} and the necessity to modify the brackets~\eqref{modified} when computing the charge algebra. It is therefore suggestive to speculate that one could embed the conformal Carrollian theory defined by $W$ into a bigger theory in which the geometry, in particular the shear $C_{ij}$, to which the theory couples becomes dynamical itself.\footnote{Related ideas have been discussed, e.g., in \cite{fioruccitalk}.} One can think of the Carrollian theory discussed here as the ``matter'' sector of a theory with dynamical, back-reacting geometry. The second equality of equation~\eqref{eq:Cdictionary} could then be interpreted as the analogue of a semi-classical Einstein equation, relating the expectation value of the matter response to the dynamical geometrical data. 
    
Assuming that both the full coupled theory and the geometric sector by itself are invariant under the BMS-symmetries, the modified brackets can be regarded as the projection of the symmetry algebra to the matter sector only. 
  
  \item[Boundary action coupled to shear.]  Clearly, to make these speculations more precise, it would be interesting to find explicit examples of Carrollian theories coupling to the shear that can be made subsequently dynamical. There exist only a small number of Carrollian actions which couple to the
    shear, see, e.g., ~\cite{Baiguera:2022lsw, Ecker:2025vnl}. A simple example of another (electric) action of that type is given by the Lagrangian
    \begin{equation}
        S
  =
  \frac12\int d^{d+1}x\,e\,\left(\delta_{ab}\mathcal{D}_0X^a\mathcal{D}_0 X^b +g\, \mathcal{D}_AX^B\mathcal{D}_BX^A\right) \, .
    \end{equation}
   Here, we introduced a Carroll matter field \(X^A=(X^0,X^a)\) that transforms as an upper
Carroll vector under Carroll transformations $\Lambda\indices{^A_B}$ and has Weyl
weight $w_X=\frac{d-1}{2}$:
\begin{align}
  \delta_\Lambda X^A
  =
  -\Lambda^A{}_B X^B
  -
  w_X z X^A  \qquad \Lambda\indices{^A_0}=0 \, .
\end{align}
The Weyl-Carroll covariant derivative is
\begin{align}
  \mathcal{D}_\mu X^A
  =
  \partial_\mu X^A
  +
  \omega_\mu{}^A{}_B X^B
  -
  w_X h_\mu X^A \, ,
  \end{align}
  with $\omega_\mu{}^A{}_B$ and $h_\mu$ connections for the Carroll-Weyl transformations. On a flat background with $h_a=0$ additionally, the boost connection $\omega\indices{^0_a}$ contains an undetermined symmetric piece $-\tfrac{1}{2}C_{ab}$ that we identify with the shear. We then find on this flat background
  \begin{equation}
 S_{\mathrm{flat}}
  =\int  d^{d+1}x\left(
  \frac12 \dot X^a\dot X_a
  +
  \frac{g}{2}
  \left[
    (\dot X^0)^2
    +
    2\dot X^a(\partial_aX^0-\frac{1}{2}C_{ab}X^b)
    +
    \partial_aX^b\partial_bX^a
  \right]\right) \, ,
    \end{equation}
    a non-zero coupling to the background shear $C_{ab}$.

This model can be used as a concrete example in which one can test the results of the present manuscript. We plan to address the exploration of such models in a future work. 
\end{description}

\section*{Acknowledgments}
We thank Glenn Barnich, Sudipta Dutta, Daniel Grumiller, Marc
Henneaux, Gerben Oling, Simon Pekar, and Matthieu Vilatte for useful
discussions.

The work of LM was supported by a Doctoral Research
Fellowship (DOC Program) granted by the OeAW. The work of JS was supported by a Postdoctoral Research Fellowship granted by the F.R.S.-FNRS (Belgium). FE, LM, and JS were supported by the Austrian Science Fund (FWF) [Grants DOI : \href{https://www.fwf.ac.at/en/research-radar/10.55776/P36619}{10.55776/P36619}, \href{https://www.fwf.ac.at/en/research-radar/10.55776/PAT1871425}{10.55776/PAT1871425}].

\appendix

\section{Consistency with the Carroll algebra}
\label{sec:cons-with-carr}

A non-trivial consistency check for the universal sector in the brackets \eqref{SET} is to be compatible with the unitary action of the Carroll algebra on tensor operators. In this appendix we work out the Carrollian analogues of expressions (4) in \cite{Deser:1967zzf}. Denoting Carroll boosts by $B_i$, spatial translations by $P_i$, spatial rotations by $J_{ij}$ and temporal translations by $H$, the Carroll algebra $\mathfrak{carr}(d+1)$ is given by
\begin{subequations}\label{eq:Carroll_algebra}
\begin{align}
    -i[P_i,B_j]&=\delta _{ij}H \\
    -i[P_i,J_{jk}]&=2\delta _{i[j}P_{k]}\\
    -i[B_i,J_{jk}]&=2\delta _{i[j}B_{k]}\\
    -i[J_{ij},J_{kl}]&=4\delta _{[i[k}J_{l]j]}
\end{align}
\end{subequations}
with all other brackets vanishing. We have decorated the brackets already with an $i$ such that the generators can be directly promoted to Hermitian operators. On the flat background \eqref{flat}, the finite-dimensional Carroll group $\text{Carr}(d+1)$ acts on spacetime coordinates as
\begin{align}
    x^0 &\to x^0 +b_ix^i +a^0 & x^i &\to R^i{}_j \, x^j +a^i.
\end{align}
We denote a group element by $(\Lambda,a)$, with the homogeneous element $\Lambda$ containing the spatial rotations $R\in \text{SO}(d)$ and Carroll boosts $b\in \mathbb{R}^d{}^\ast$, whereas $a$ are spacetime translations. $\Lambda$ admits the matrix representation
\begin{align}
\Lambda^{\mu }{}_\nu =
    \begin{pmatrix}
        1 & b_i \\
        0 & R^i{}_j
    \end{pmatrix} \approx \begin{pmatrix}
        1 & 0 \\
        0 & \delta^i_j 
    \end{pmatrix} + \begin{pmatrix}
        0 & \lambda _i \\
        0 & \omega ^i{}_j
    \end{pmatrix}
\end{align}
where we also considered an infinitesimal expansion as $\Lambda^\mu{}_\nu \approx \delta^\mu_\nu + \omega^\mu{}_\nu$ and $a^\mu \approx \epsilon^\mu$. 
We construct a unitary operator 
\begin{align}
    U(\Lambda,a)=\exp \Big(i\epsilon ^0 H  +i\epsilon^i P_i+i\lambda_i B^i
    -\frac{i}{2}\omega _{ij}J^{ij}\Big)
\end{align}
where $\{J_{ij},B_i,P_i,H\}$ are the algebra elements promoted to Hermitian operators. Demanding this to be a representation of the Carroll group,
$U(\Lambda,a)U(\Bar{\Lambda},\Bar{a})=U(\Lambda\Bar{\Lambda},\Lambda\Bar{a}+a)$, one finds
\begin{align}
    U(\Lambda,a)U(1+\omega ,\epsilon )U^{-1}(\Lambda,a)=U(1+\Lambda\omega \Lambda^{-1},\Lambda\epsilon -\Lambda\omega
    \Lambda^{-1}a) \, ,
\end{align}
from which the algebra \eqref{eq:Carroll_algebra} is recovered by infinitesimally expanding both sides.

Let us now see how these transformations act on fields. For a $(1,1)$-tensor $\mathbf{T}^\mu{}_\nu$ we have on general grounds
\begin{align}
     U(\Lambda,a)\mathbf{T}^\mu {}_\nu (x)U^{-1}(\Lambda,a)=(\Lambda^{-1})^\mu {}_\alpha
     \Lambda^\beta {}_\nu \mathbf{T}^\alpha {}_\beta (\Lambda x+a)
\end{align}
and expanding both sides to linear order in the various infinitesimal parameters one finds that translations act without mixing tensor components, $i[P_i,\mathbf{T}^\mu{}_\nu]=\partial_i\mathbf{T}^\mu{}_\nu$ and $i[H,\mathbf{T}^\mu{}_\nu]=\partial_0
\mathbf{T}^\mu{}_\nu$, while rotations and boosts give
\begin{align}\label{consistency}
    i[J^{ij},\mathbf{T}^0{}_0]&=2x^{[i} \partial^{j]}\mathbf{T}^0{}_0
    & &i[B^i,\mathbf{T}^0{}_0]=x^i\partial _0\mathbf{T}^0{}_0-\mathbf{T}^i{}_0\\
    \nonumber
    i[J^{ij},\mathbf{T}^0{}_k]&=2x^{[i} \partial^{j]}\mathbf{T}^0{}_k
    -2\delta ^{m[i}\delta ^{j]}_k\mathbf{T}^0{}_m
    & &i[B^i,\mathbf{T}^0{}_k]=x^i\partial _0\mathbf{T}^0{}_k-\mathbf{T}^i{}_k
    +\delta ^i_k\mathbf{T}^0{}_0\\
    \nonumber
    i[J^{ij},\mathbf{T}^k{}_m]&=2x^{[i} \partial^{j]}\mathbf{T}^k{}_m
    -2\delta ^{l[i}\delta ^{j]}_m\mathbf{T}^k{}_l+2\delta ^{k[i}\mathbf{T}^{j]}{}_m
    & &i[B^i,\mathbf{T}^k{}_m]=x^i\partial _0\mathbf{T}^k{}_m
    +\delta _m^i \mathbf{T}^k{}_0
\end{align}
The remaining set, involving $\mathbf{T}^k{}_0$, can be disregarded: the boost Ward identity sets $ \mathbf{T}^i{}_0=0$, which is consistent with the right hand sides above without imposing any further constraint. We define charges for a Killing vector $\xi$ as in the main text
\begin{equation}
  Q_\xi = -\int d^dx\, \xi^\mu \mathbf{T}^0{}_\mu \,.
\end{equation}
Now, since the spacetime symmetry generators are canonically given by integrals over constant time hypersurfaces,
\begin{align}
    H&= Q_{\partial_0} = -\int d ^dx\, \mathbf{T}^0{}_0(x) & P_{i}&=Q_{\partial_i} =-\int d^d x\, \mathbf{T}^0{}_i(x)\\
    \nonumber
    B_i&=Q_{x_i\partial_0} =-\int d^dx\, (x_i \mathbf{T}^0{}_0(x)) & J_{ij}&=Q_{2 x_{[i}\partial_{j]}} =-\int d^d x\, (x_i\mathbf{T}^0{}_j(x)-x_j\mathbf{T}^0{}_i(x) )
\end{align}
the consistency of the universal brackets \eqref{SET} with \eqref{consistency} can only be achieved by the vanishing of the zeroth and first moments of certain Schwinger terms:
\begin{align}
  \label{moments}
    \int d^{d}x\, \tau ^0 {}_\nu {}^\alpha {}_\beta (x^k,y^k)&=0, & \int d^{d}x\, x^k  \tau^0 {}_0 {}^\alpha {}_\beta (x^k,y^k)&=0 \, 
\end{align}
together with 
\begin{equation}\label{moments2}
\int d^{d}x\, \left[ x^j  \tau^0 {}_i{}^\alpha {}_\beta (x^k,y^k) - x^i  \tau^0 {}_j{}^\alpha {}_\beta (x^k,y^k) \right]=0 \, .
\end{equation}
This in turn implies that for the quantum EMT to transform covariantly under the finite Carroll group, the admitted Schwinger terms would have to carry a specific derivative structure on the deltas to fulfill \eqref{moments} and \eqref{moments2}. In particular, the Weyl conditions \eqref{weylcondition} and the Schwinger terms derived in section \ref{sec:boost-anomaly} satisfy these constraints.

\bibliographystyle{utphys} %~/texmf/bibtex/bst/
\bibliography{bibl} %~/texmf/bibtex/bib/

@article{Cotler:2025npu,
    author = "Cotler, Jordan and Dhivakar, Prateksh and Jensen, Kristan",
    title = "{Carrollian Holographic Duals Are Nonlocal}",
    eprint = "2512.05072",
    archivePrefix = "arXiv",
    primaryClass = "hep-th",
    doi = "10.1103/f26s-4p77",
    journal = "Phys. Rev. Lett.",
    volume = "137",
    number = "9",
    pages = "091601",
    year = "2026"
}

@article{Chicherin:2025keq,
    author = "Chicherin, Dmitry and Korchemsky, Gregory P. and Sokatchev, Emery and Zhiboedov, Alexander",
    title = "{Energy correlators in four-dimensional gravity}",
    eprint = "2512.23791",
    archivePrefix = "arXiv",
    primaryClass = "hep-th",
    reportNumber = "CERN-TH-2025-266, LAPTH-060/25",
    doi = "10.1007/JHEP07(2026)275",
    journal = "JHEP",
    volume = "07",
    pages = "275",
    year = "2026"
}

@article{Gonzalez:2025ene,
    author = "Gonz{\'a}lez, Hern{\'a}n A. and Salzer, Jakob",
    title = "{Energy detectors and asymptotic symmetries}",
    eprint = "2510.27348",
    archivePrefix = "arXiv",
    primaryClass = "hep-th",
    doi = "10.1007/JHEP07(2026)241",
    journal = "JHEP",
    volume = "07",
    pages = "241",
    year = "2026"
}

@article{Himwich:2026exq,
    author = "Himwich, Elizabeth and Pate, Monica",
    title = "{Infinite Symmetry Algebras in Four-Dimensional Conformal Field Theories}",
    eprint = "2607.28718",
    archivePrefix = "arXiv",
    primaryClass = "hep-th",
    month = "7",
    year = "2026"
}

@article{Moult:2025njc,
    author = "Moult, Ian and Narayanan, Sruthi A. and Pasterski, Sabrina",
    title = "{Memory correlators and Ward identities in the 'in-in' formalism}",
    eprint = "2512.02825",
    archivePrefix = "arXiv",
    primaryClass = "hep-th",
    doi = "10.1007/JHEP09(2026)239",
    journal = "JHEP",
    volume = "09",
    pages = "239",
    year = "2026"
}

@conference{fioruccitalk,
    author =
"A.~Fiorucci",
    title = "Hamiltonian Mechanics with External Sources \emph{(based on unpublished results with G. Barnich and R. Ruzziconi)}",
    year = "Talk at 2nd
  Carroll Workshop, Mons, September 2023"
}

@article{Gross:1969zow,
    author = "Gross, D. J. and Jackiw, R.",
    title = "{Construction of covariant and gauge invariant t* products}",
    doi = "10.1016/0550-3213(69)90207-7",
    journal = "Nucl. Phys. B",
    volume = "14",
    pages = "269--284",
    year = "1969"
}

@article{Grumiller:2025rtm,
    author = "Grumiller, Daniel and Mele, Lea and Montecchio, Luciano",
    title = "{Carroll spinors}",
    eprint = "2509.19426",
    archivePrefix = "arXiv",
    primaryClass = "hep-th",
    reportNumber = "TUW-25-05",
    doi = "10.1142/s0217732325400036",
    journal = "Mod. Phys. Lett. A",
    volume = "41",
    number = "27",
    pages = "2540003",
    year = "2026"
}

@article{Hartong:2026bvh,
    author = "Hartong, Jelle",
    title = "{The Geometry of Gravitational Radiation}",
    eprint = "2609.09954",
    archivePrefix = "arXiv",
    primaryClass = "hep-th",
    month = "9",
    year = "2026"
}

@article{Hartong:2025jpp,
    author = "Hartong, Jelle and Have, Emil and Nenmeli, Vijay and Oling, Gerben",
    title = "{Boundary Energy-Momentum Tensors for Asymptotically Flat Spacetimes}",
    eprint = "2505.05432",
    archivePrefix = "arXiv",
    primaryClass = "hep-th",
    month = "5",
    year = "2025"
}

@article{Hartong:2026rbr,
    author = "Hartong, Jelle and Have, Emil and Nenmeli, Vijay and Oling, Gerben",
    title = "{The Energy-Momentum-News Complex near Future Null Infinity}",
    eprint = "2607.07872",
    archivePrefix = "arXiv",
    primaryClass = "hep-th",
    month = "7",
    year = "2026"
}

@article{Fiorucci:2025twa,
    author = "Fiorucci, Adrien and Pekar, Simon and Marios Petropoulos, P. and Vilatte, Matthieu",
    title = "{Carrollian-Holographic Derivation of Gravitational Flux-Balance Laws}",
    eprint = "2505.00077",
    archivePrefix = "arXiv",
    primaryClass = "hep-th",
    reportNumber = "CPHT-RR035.042025",
    doi = "10.1103/qv17-ks32",
    journal = "Phys. Rev. Lett.",
    volume = "135",
    number = "26",
    pages = "261602",
    year = "2025"
}

@article{Krishna:2026sba,
    author = "Krishna, Hare and Patil, Vaishnavi",
    title = "{Aspects of Carrollian field theory from holography}",
    eprint = "2608.12241",
    archivePrefix = "arXiv",
    primaryClass = "hep-th",
    reportNumber = "UT-WI-22-2026",
    month = "8",
    year = "2026"
}

@article{Campoleoni:2022wmf,
    author = "Campoleoni, Andrea and Ciambelli, Luca and Delfante, Arnaud and Marteau, Charles and Petropoulos, P. Marios and Ruzziconi, Romain",
    title = "{Holographic Lorentz and Carroll frames}",
    eprint = "2208.07575",
    archivePrefix = "arXiv",
    primaryClass = "hep-th",
    doi = "10.1007/JHEP12(2022)007",
    journal = "JHEP",
    volume = "12",
    pages = "007",
    year = "2022"
}

@article{Baiguera:2022lsw,
    author = "Baiguera, Stefano and Oling, Gerben and Sybesma, Watse and S{\o}gaard, Benjamin T.",
    title = "{Conformal Carroll scalars with boosts}",
    eprint = "2207.03468",
    archivePrefix = "arXiv",
    primaryClass = "hep-th",
    reportNumber = "NORDITA 2022-047",
    doi = "10.21468/SciPostPhys.14.4.086",
    journal = "SciPost Phys.",
    volume = "14",
    number = "4",
    pages = "086",
    year = "2023"
}

@inproceedings{Geroch:1977big,
    author = "Geroch, Robert",
    title = "{Asymptotic Structure of Space-Time}",
    booktitle = "{Symposium on Asymptotic Structure of Space-Time}",
    doi = "10.1007/978-1-4684-2343-3_1",
    year = "1977"
}

@article{Deser:1967zzf,
    author = "Deser, Stanley and Boulware, D.",
    title = "{Stress-Tensor Commutators and Schwinger Terms}",
    doi = "10.1063/1.1705368",
    journal = "J. Math. Phys.",
    volume = "8",
    pages = "1468",
    year = "1967"
}

@article{Fredenhagen:2026pia,
    author = "Fredenhagen, Stefan and Prohazka, Stefan and Tiefenbacher, Robert",
    title = "{Carrollian quantum states and flat space holography}",
    eprint = "2604.22745",
    archivePrefix = "arXiv",
    primaryClass = "hep-th",
    doi = "10.1007/JHEP11(2026)018",
    journal = "JHEP",
    volume = "11",
    pages = "018",
    year = "2026"
}

@article{Cotler:2024cia,
    author = "Cotler, Jordan and Jensen, Kristan and Prohazka, Stefan and Riegler, Max and Salzer, Jakob",
    title = "{Soft gravitons in three dimensions}",
    eprint = "2411.13633",
    archivePrefix = "arXiv",
    primaryClass = "hep-th",
    doi = "10.1007/JHEP07(2025)002",
    journal = "JHEP",
    volume = "07",
    pages = "002",
    year = "2025"
}

@article{Cotler:2024xhb,
    author = "Cotler, Jordan and Jensen, Kristan and Prohazka, Stefan and Raz, Amir and Riegler, Max and Salzer, Jakob",
    title = "{Quantizing Carrollian field theories}",
    eprint = "2407.11971",
    archivePrefix = "arXiv",
    primaryClass = "hep-th",
    doi = "10.1007/JHEP10(2024)049",
    journal = "JHEP",
    volume = "10",
    pages = "049",
    year = "2024"
}

@phdthesis{Ecker:2025vnl,
    author = "Ecker, Florian",
    title = "{Carroll symmetries in field theory and gravity}",
    eprint = "2603.12902",
    archivePrefix = "arXiv",
    primaryClass = "hep-th",
    doi = "10.34726/hss.2026.113247",
    school = "Vienna, Tech. U., TU. Vienna",
    month = "12",
    year = "2025"
}

@article{Ashtekar:1981hw,
    author = "Ashtekar, A.",
    title = "{Radiative Degrees of Freedom of the Gravitational Field in Exact General Relativity}",
    doi = "10.1063/1.525169",
    journal = "J. Math. Phys.",
    volume = "22",
    pages = "2885--2895",
    year = "1981"
}

@article{Sachs:1962wk,
    author = "Sachs, R. K.",
    title = "{Gravitational waves in general relativity. 8. Waves in asymptotically flat space-times}",
    doi = "10.1098/rspa.1962.0206",
    journal = "Proc. Roy. Soc. Lond. A",
    volume = "270",
    pages = "103--126",
    year = "1962"
}

@article{deBoer:2023fnj,
    author = "de Boer, Jan and Hartong, Jelle and Obers, Niels A. and Sybesma, Watse and Vandoren, Stefan",
    title = "{Carroll stories}",
    eprint = "2307.06827",
    archivePrefix = "arXiv",
    primaryClass = "hep-th",
    reportNumber = "NORDITA-2023-036",
    doi = "10.1007/JHEP09(2023)148",
    journal = "JHEP",
    volume = "09",
    pages = "148",
    year = "2023"
}

@article{Henneaux:2021yzg,
    author = "Henneaux, Marc and Salgado-Rebolledo, Patricio",
    title = "{Carroll contractions of Lorentz-invariant theories}",
    eprint = "2109.06708",
    archivePrefix = "arXiv",
    primaryClass = "hep-th",
    doi = "10.1007/JHEP11(2021)180",
    journal = "JHEP",
    volume = "11",
    pages = "180",
    year = "2021"
}

@article{Donnay:2022wvx,
    author = "Donnay, Laura and Fiorucci, Adrien and Herfray, Yannick and Ruzziconi, Romain",
    title = "{Bridging Carrollian and celestial holography}",
    eprint = "2212.12553",
    archivePrefix = "arXiv",
    primaryClass = "hep-th",
    doi = "10.1103/PhysRevD.107.126027",
    journal = "Phys. Rev. D",
    volume = "107",
    number = "12",
    pages = "126027",
    year = "2023"
}

@article{Ciambelli:2019lap,
    author = "Ciambelli, Luca and Leigh, Robert G. and Marteau, Charles and Petropoulos, P. Marios",
    title = "{Carroll Structures, Null Geometry and Conformal Isometries}",
    eprint = "1905.02221",
    archivePrefix = "arXiv",
    primaryClass = "hep-th",
    reportNumber = "CPHT-RR025.052019, CPHT-RR010.022019",
    doi = "10.1103/PhysRevD.100.046010",
    journal = "Phys. Rev. D",
    volume = "100",
    number = "4",
    pages = "046010",
    year = "2019"
}

@InProceedings{10.1007/3-540-09238-2_90,
author="Teitelboim, Claudio",
editor="Beiglb{\"o}ck, Wolf
and B{\"o}hm, Arno
and Takasugi, E.",
title="Surface deformations, their square root and the signature of spacetime",
booktitle="Group Theoretical Methods in Physics",
year="1979",
publisher="Springer Berlin Heidelberg",
address="Berlin, Heidelberg",
pages="362--371",
isbn="978-3-540-35345-4"
}

@article{Nguyen:2020hot,
    author = "Nguyen, K\'evin and Salzer, Jakob",
    title = "{The Effective Action of Superrotation Modes}",
    eprint = "2008.03321",
    archivePrefix = "arXiv",
    primaryClass = "hep-th",
    month = "8",
    year = "2020"
}

@article{Teitelboim:1972vw,
      author         = "Teitelboim, Claudio",
      title          = "{How commutators of constraints reflect the space-time
                        structure}",
      journal        = "Annals Phys.",
      volume         = "79",
      year           = "1973",
      pages          = "542-557",
      doi            = "10.1016/0003-4916(73)90096-1",
      SLACcitation   = "%%CITATION = APNYA,79,542;%%"
}

@article{RevModPhys.34.592,
  title = {The Conditions for a Quantum Field Theory to be Relativistic},
  author = {Dirac, P. A. M.},
  journal = {Rev. Mod. Phys.},
  volume = {34},
  issue = {4},
  pages = {592--596},
  numpages = {0},
  year = {1962},
  month = {Oct},
  publisher = {American Physical Society},
  doi = {10.1103/RevModPhys.34.592}
}

@article{Cordova:2018ygx,
      author         = "Córdova, Clay and Shao, Shu-Heng",
      title          = "{Light-ray Operators and the BMS Algebra}",
      journal        = "Phys. Rev.",
      volume         = "D98",
      year           = "2018",
      number         = "12",
      pages          = "125015",
      doi            = "10.1103/PhysRevD.98.125015",
      eprint         = "1810.05706",
      archivePrefix  = "arXiv",
      primaryClass   = "hep-th",
      SLACcitation   = "%%CITATION = ARXIV:1810.05706;%%"
}

@article{Schwinger:1963zza,
      author         = "Schwinger, Julian",
      title          = "{Commutation Relations and Conservation Laws}",
      journal        = "Phys. Rev.",
      volume         = "130",
      year           = "1963",
      pages          = "406-409",
      doi            = "10.1103/PhysRev.130.406",
      SLACcitation   = "%%CITATION = PHRVA,130,406;%%"
}

@article{Henneaux:1979vn,
      author         = "Henneaux, Marc",
      title          = "{Geometry of Zero Signature Space-times}",
      journal        = "Bull. Soc. Math. Belg.",
      volume         = "31",
      year           = "1979",
      pages          = "47-63",
      reportNumber   = "PRINT-79-0606 (PRINCETON)",
      SLACcitation   = "%%CITATION = BMBEA,31,47;%%"
}

@article{Duval:2014uva,
      author         = "Duval, C. and Gibbons, G. W. and Horvathy, P. A.",
      title          = "{Conformal Carroll groups and BMS symmetry}",
      journal        = "Class. Quant. Grav.",
      volume         = 31,
      year           = 2014,
      pages          = 092001,
      doi            = "10.1088/0264-9381/31/9/092001",
      eprint         = "1402.5894",
      archivePrefix  = "arXiv",
      primaryClass   = "gr-qc",
      SLACcitation   = "%%CITATION = ARXIV:1402.5894;%%"
}

@article{Hartong:2015xda,
      author         = "Hartong, Jelle",
      title          = "{Gauging the Carroll Algebra and Ultra-Relativistic
                        Gravity}",
      journal        = "JHEP",
      volume         = "08",
      year           = "2015",
      pages          = "069",
      doi            = "10.1007/JHEP08(2015)069",
      eprint         = "1505.05011",
      archivePrefix  = "arXiv",
      primaryClass   = "hep-th",
      SLACcitation   = "%%CITATION = ARXIV:1505.05011;%%"
}

@article{Barnich:2006av,
      author         = "Barnich, Glenn and Compere, Geoffrey",
      title          = "{Classical central extension for asymptotic symmetries at
                        null infinity in three spacetime dimensions}",
      journal        = "Class. Quant. Grav.",
      volume         = "24",
      year           = "2007",
      pages          = "F15-F23",
      doi            = "10.1088/0264-9381/24/5/F01, 10.1088/0264-9381/24/11/C01",
      eprint         = "gr-qc/0610130",
      archivePrefix  = "arXiv",
      primaryClass   = "gr-qc",
      reportNumber   = "ULB-TH-06-08",
      SLACcitation   = "%%CITATION = GR-QC/0610130;%%"
}

@article{Ashtekar:2014zsa,
      author         = "Ashtekar, Abhay",
      title          = "{Geometry and Physics of Null Infinity}",
      year           = "2014",
      eprint         = "1409.1800",
      archivePrefix  = "arXiv",
      primaryClass   = "gr-qc",
      reportNumber   = "IGC-14-9-1",
      SLACcitation   = "%%CITATION = ARXIV:1409.1800;%%"
}

@article{Bondi:1962px,
      author         = "Bondi, H. and van der Burg, M. G. J. and Metzner, A. W.
                        K.",
      title          = "{Gravitational waves in general relativity. 7. Waves from
                        axisymmetric isolated systems}",
      journal        = "Proc. Roy. Soc. Lond.",
      volume         = "A269",
      year           = "1962",
      pages          = "21-52",
      doi            = "10.1098/rspa.1962.0161",
      SLACcitation   = "%%CITATION = PRSLA,A269,21;%%"
}

@article{Barnich:2011mi,
      author         = "Barnich, Glenn and Troessaert, Cedric",
      title          = "{BMS charge algebra}",
      journal        = "JHEP",
      volume         = "12",
      year           = "2011",
      pages          = "105",
      doi            = "10.1007/JHEP12(2011)105",
      eprint         = "1106.0213",
      archivePrefix  = "arXiv",
      primaryClass   = "hep-th",
      reportNumber   = "ULB-TH-11-10",
      SLACcitation   = "%%CITATION = ARXIV:1106.0213;%%"
}

@Article{Bagchi:2012yk,
  Title                    = {{Flat-Space Chiral Gravity}},
  Author                   = {Bagchi, Arjun and Detournay, Stephane and Grumiller, Daniel},
  Journal                  = {Phys.Rev.Lett.},
  Year                     = {2012},
  Pages                    = {151301},
  Volume                   = {109},

  Archiveprefix            = {arXiv},
  Doi                      = {10.1103/PhysRevLett.109.151301},
  Eprint                   = {1208.1658},
  Primaryclass             = {hep-th},
  Slaccitation             = {%%CITATION = ARXIV:1208.1658;%%}
}

@Article{Gonzo2021,
  author        = {Gonzo, Riccardo and Pokraka, Andrzej},
  journal       = {JHEP},
  title         = {{Light-ray operators, detectors and gravitational event shapes}},
  year          = {2021},
  pages         = {015},
  volume        = {05},
  archiveprefix = {arXiv},
  doi           = {10.1007/JHEP05(2021)015},
  eprint        = {2012.01406},
  primaryclass  = {hep-th},
  reportnumber  = {SAGEX-20-24-E},
}

@article{Delfante:2026ads,
    author = "Delfante, Arnaud and Markou, Chrysoula",
    title = "{Carroll-Weyl symmetries in flat holography}",
    eprint = "2609.36026",
    archivePrefix = "arXiv",
    primaryClass = "hep-th",
    month = "9",
    year = "2026"
}

@article{Schwinger:1959xd,
    author = "Schwinger, Julian S.",
    editor = "Milton, K. A.",
    title = "{Field theory commutators}",
    doi = "10.1103/PhysRevLett.3.296",
    journal = "Phys. Rev. Lett.",
    volume = "3",
    pages = "296--297",
    year = "1959"
}

@article{Barnich:2014kra,
    author = "Barnich, Glenn and Oblak, Blagoje",
    title = "{Notes on the BMS group in three dimensions: I. Induced representations}",
    eprint = "1403.5803",
    archivePrefix = "arXiv",
    primaryClass = "hep-th",
    doi = "10.1007/JHEP06(2014)129",
    journal = "JHEP",
    volume = "06",
    pages = "129",
    year = "2014"
}

@article{Bagchi:2021gai,
    author = "Bagchi, Arjun and Dutta, Sudipta and Kolekar, Kedar S. and Sharma, Punit",
    title = "{BMS field theories and Weyl anomaly}",
    eprint = "2104.10405",
    archivePrefix = "arXiv",
    primaryClass = "hep-th",
    doi = "10.1007/JHEP07(2021)101",
    journal = "JHEP",
    volume = "07",
    pages = "101",
    year = "2021"
}

@article{Kraus:2005zm,
    author = "Kraus, Per and Larsen, Finn",
    title = "{Holographic gravitational anomalies}",
    eprint = "hep-th/0508218",
    archivePrefix = "arXiv",
    doi = "10.1088/1126-6708/2006/01/022",
    journal = "JHEP",
    volume = "01",
    pages = "022",
    year = "2006"
}

@incollection{Jackiw:1985,
    author    = "Jackiw, R.",
    title     = "{Field Theoretic Investigations in Current Algebra}",
    booktitle = "{Current Algebra and Anomalies}",
    publisher = "Princeton University Press",
    year      = "1985",
    pages     = "81--210"
}

\end{document}